\documentclass[aps,prd,preprintnumbers,showpacs,showkeys,nofootinbib,
superscriptaddress,fleqn,floatfix,tightenlines,10pt]{revtex4-1}
\usepackage{amsmath,amsfonts,amssymb,amscd,amsxtra,amsthm}
\usepackage{graphicx}  
\usepackage{epstopdf}
\usepackage{dcolumn}  
\usepackage{bm}          
\usepackage{slashed}
\usepackage{cancel}
\usepackage{float}
\usepackage{mathtools} 
\usepackage{amsbsy}
\usepackage{amstext}

\usepackage[utf8]{inputenc} 
\usepackage{booktabs}
\usepackage[normalem]{ulem} 
\usepackage[dvipsnames]{xcolor} 
\usepackage{tabularx}
\usepackage{enumitem}  
\usepackage{array} 
\usepackage{slashed}
\usepackage{tikz}
\usepackage{float}
\usepackage{multirow}
\renewcommand\sout{\bgroup \color{red} \ULdepth=-.5ex \ULset}

\makeatletter

\begin{document}  
\preprint{INHA-NTG-05/2020}
\title{Electric quadrupole form factors of singly heavy baryons with
  spin 3/2}   
\author{June-Young Kim}
\email[E-mail: ]{Jun-Young.Kim@ruhr-uni-bochum.de}
\affiliation{Institut f\"ur Theoretische Physik II, Ruhr-Universit\"at
  Bochum, D-44780 Bochum, Germany}
\affiliation{Department of Physics, Inha University, Incheon 22212,
Republic of Korea}
\author{Hyun-Chul Kim}
\email[E-mail: ]{hchkim@inha.ac.kr}
\affiliation{Department of Physics, Inha University, Incheon 22212,
Republic of Korea}
\affiliation{School of Physics, Korea Institute for Advanced Study 
  (KIAS), Seoul 02455, Republic of Korea}
\date{\today}
\begin{abstract}
We study the electromagnetic form factors of the lowest-lying singly
heavy baryons in a pion mean-field approach, which is also known as
the SU(3) chiral quark-soliton model. In the limit of the heavy-quark
mass, the dynamics inside a singly heavy baryon is governed by the
$N_c-1$ valence quarks, while the heavy quark remains as a mere static
color source. In this framework, a singly heavy baryon is described by
combining the colored soliton with the singly heavy quark. In the
infinitely heavy-quark mass limit, we can compute the electric
quadrupole form factors of the baryon sextet with spin 3/2 with the
rotational $1/N_c$ and linear corrections of the explicit flavor SU(3)
symmetry breaking taken into account. We find that the sea-quark
contributions or the Dirac-sea level contributions dominate over the
valence-quark contributions in lower $Q^2$ region. We examined the
effects of explicit flavor SU(3) symmetry breaking in detail. The
numerical results are also compared with the recent data from the
lattice calculation with the unphysical value of the pion mass
considered, which was used in the lattice calculation. 
\end{abstract}
\maketitle
\section{Introduction}
Conventional lowest-lying singly heavy baryons consist of a heavy
quark and two light valence quarks. In the limit of the
infinitely heavy-quark mass ($m_Q\to \infty$), the physics of singly
heavy baryons becomes simple: The spin of the heavy quark $\bm{J}_Q$
is conserved in this limit and hence it leads also to the conservation
of the spin of the light-quark degrees of freedom,
i.e. $\bm{J}_L=\bm{J}-\bm{J}_Q$. This is known as the heavy-quark spin
symmetry~\cite{Isgur:1989vq,Georgi:1990um}. In the $m_Q\to \infty$
limit, we do not distinguish a charm quark from a bottom quark, which
gives up heavy-quark flavor symmetry. On the other hand, chiral
symmetry and its spontaneous breakdown still play an important part
in describing the singly heavy baryons because of the presence
of the light quarks inside a singly heavy baryon~\cite{Yan:1992gz}. 
Then the singly heavy baryons consisting of two light valence quarks
can be represented in terms of irreducible representations of flavor
SU(3) symmetry: $\bm{3}\otimes \bm{3}=\bar{\bm{3}}\oplus \bm{6}$, thus
we have the two representations for the lowest-lying singly heavy
baryons, i.e. the baryon antitriplet and sextet. The baryon
antitriplet has the total spin $J=1/2$ that comes from $J_Q=1/2$ and
$J_L=0$, whereas the baryon sextet can have either $J=1/2$ or $J=3/2$
with $J_L=1$ and $J_Q=1/2$. 

In a pion mean-field approach, which is also known as the SU(3) chiral
quark-soliton model ($\chi$QSM), a singly heavy baryon can be viewed as the
$N_c-1$ valence quarks bound by the pion mean fields that are created
from the presence of the $N_c-1$ valence quarks~\cite{Diakonov:2010tf,
Yang:2016qdz}. In fact, this idea is taken from Witten's seminal paper
on baryons in the large $N_c$ limit~\cite{Witten:1979kh}. This pion
mean-field approach has successfully reproduced the mass spectra of
the lowest-lying singly heavy baryons~\cite{Yang:2016qdz} and even
explained the nontrivial isospin mass splittings of
them~\cite{Yang:2020klp}. Interestingly, the corrections from the
heavy quark mass are indeed negligible in the description of the
isospin mass splittings of the singly heavy baryons as shown in
Ref.~\cite{Yang:2020klp}, although they provide hyperfine interactions 
to remove the spin degeneracy of the baryon sextet. 

Recently, the electromagnetic (EM) form factors of singly heavy baryons
have been studied for the first time within lattice
QCD~\cite{Can:2013tna,Can:2015exa}. While there are no experimental
data on the EM form factors of the singly heavy baryons to date, the
results from the lattice calculation provide a clue to the internal
structure of singly heavy baryons. Thus, it is also of great
importance to investigate the EM form factors of the singly heavy
baryons. In Ref.~\cite{Kim:2018nqf,Kim:2019wbg}, we have studied the
electric monopole and magnetic dipole form factors of the singly heavy
baryons in detail, based on the $\chi$QSM. Since we consider the limit
of the infinitely heavy-quark mass, there is no physical difference
between the heavy baryons with spin 1/2 and those with 3/2 except for
the value of the spin. On the other hand, the baryon sextet with spin
3/2 has yet another structure that arises from its higher spin, which
is revealed by the electric quadrupole ($E2$) form factor.  
The $E2$ form factor of a baryon exhibits how the baryon is
deformed. It is also known that the pion clouds play a significant 
role in understanding this deformation~\cite{Pascalutsa:2006up}. This
will be also discussed in the present work.  We will also examine the
effects of flavor SU(3) symmetry breaking on the $E2$ form factors of
the baryon sextet with spin 3/2. The numerical results for
$\Omega_c^{*0}$ will be compared with that from the lattice
calculation. 

The present work is organized as follows: In Section II, we briefly 
recapitulate the general formalism for the electric quadrupole form
factors within the framework of the chiral quark-soliton model. In
Section III, we present the numerical results and discuss them in
detail. The final Section is devoted to the summary and conclusion. 
\section{Electric quadrupole form factors in the $\chi$QSM}
We start with the EM current for a singly heavy baryon, which is
defined by  
\begin{align}
  \label{eq:LHcurrent}
J^\mu (x) = \bar{\psi} (x) \gamma^\mu \hat{\mathcal{Q}} \psi(x) + e_{Q}
  \bar{\Psi}   \gamma^\mu \Psi,    
\end{align}
where $\psi(x)$ stands for the light-quark field $\psi=(u,d,s)$ in
SU(3) flavor space and $\Psi$ denotes the heavy-quark field for the
charmed or bottom quark. The charge operator $\mathcal{Q}$ is
expressed as 
\begin{align}
\hat{\mathcal{Q}} =
  \begin{pmatrix}
\frac23 & 0 & 0 \\
0 & -\frac13 & 0 \\    
0 & 0 & -\frac13
  \end{pmatrix}
=\frac12\left(\lambda_3 + \frac1{\sqrt{3}} \lambda_8\right). 
\end{align}
The $e_Q$ in the second term in Eq.~\eqref{eq:LHcurrent} denotes the
charge corresponding to a heavy quark, which has the value $2/3$ for
the charm quark and $-1/3$ for the bottom quark. The matrix element of
$J^\mu$ between baryons with spin 3/2 can be parametrized in terms of
four different real form factors as follows:  
\begin{align}
\langle B(p',s) | J^{\mu}(0) | B(p,s) \rangle 
&= - \overline{u}^{\alpha}(p',s) \left[ \gamma^{\mu} \left \{
  F^{B}_{1}(q^2) \eta_{\alpha \beta} + F^{B}_{3}(q^2) \frac{ q_{\alpha} q_{\beta}
  }{4M_{B}^{2}}  \right \}\right. \cr
&\hspace{0.4cm} \left.  + \; i\frac{\sigma^{\mu \nu} q_{\nu}}{2M_{B}}
  \left \{ F^{B}_{2}(q^2) \eta_{\alpha \beta} + F^{B}_{4} (q^2)\frac{q_{\alpha}
  q_{\beta}}{4 M_{B}^2}  \right \}  \right ]{u}^{\beta}(p,s), 
\label{eq:MatrixEl1}
\end{align}
where $M_B$ denotes the mass of a singly heavy baryon in the baryon
sextet with spin 3/2. The metric tensor
$\eta_{\alpha\beta}$ of Minkowski space is defined as
$\eta_{\alpha\beta} =\mathrm{diag}(1,\,-1,\,-1,\,-1)$. $q_\alpha$
represents the momentum transfer $q_\alpha=p'_\alpha-p_\alpha$ and its
square is written as $q^2=-Q^2$ with $Q^2 >0$. $u^\alpha (p,\,s)$
means the Rarita-Schwinger spinor for a singly heavy baryon
with spin 3/2, carrying the momentum $p$ and the spin component $s$
projected along the direction of the momentum. $\sigma^{\mu\nu}$
designates the antisymmetric tensor   
$\sigma^{\mu\nu}=i[\gamma^\mu,\,\gamma^\nu]/2$. 
Note that when one takes the limit of the 
infinitely heavy quark mass ($m_Q\to \infty$), the heavy-quark current
given in the second part of Eq.~\eqref{eq:LHcurrent} can be safely
neglected for the EM form factors. It gives only a constant contribution
to the electric form factors as already shown in
Ref.~\cite{Kim:2018nqf}.   

It is more convenient to introduce the Sachs-type form factors or the
multipole EM form factors, in particular, when the EM structure of a
baryon with spin 3/2 is examined. The electric quadrupole form factor
reveals how the shape of a baryon with spin 3/2 is deviated from the
rotationally symmetric one. The Sachs-type form factors can be
expressed in terms of $F_i^B$ given in Eq.~\eqref{eq:MatrixEl1} 
\begin{align}
G_{E0}^B (Q^2) &= \left(1+\frac23 \tau\right) [F_1^B(Q^2) - \tau
                 F_2^B(Q^2)] -\frac13 \tau(1+\tau) [F_3^B(Q^2) - \tau
                 F_4^B(Q^2)],\cr
G_{E{2}}^B (Q^{2}) &= [F_{1}(Q^2)-\tau F_{2}(Q^2)]- \frac{1}{2} (1+ \tau)
  [F_{3}(Q^2) - \tau F_{4}(Q^2)],\cr
G_{M1}^B(Q^2) &= \left(1+ \frac45 \tau\right) [F_1^B(Q^2) +
                F_2^B(Q^2)] -\frac25 \tau(1+\tau) [F_3^B(Q^2) +
                F_4^B(Q^2)],  \cr
G_{M3}^B(Q^2) &= [F_1^B(Q^2) + F_2^B(Q^2)] - \frac12 (1+\tau)
                [F_3^B(Q^2) + F_4^B(Q^2)],
\end{align}
where $\tau=Q^2/4M_B^2$. Since $G_{E0}^B$, $G_{M1}^B$ have been
already investigated in Ref.~\cite{Kim:2018nqf}, we will focus on the
electric quadrupole form factors of the baryon sextet with spin 3/2,
i.e., $G_{E2}^B$ in the present work.  
At $Q^2=0$, $G_{E2}(0)$ yields the electric quadrupole moment  
\begin{align}
\mathcal{Q}_B &= \frac{e}{M_B^2} G_{E2}^B(0) = \frac{e}{M_B^2} \left[e_B
      -\frac12       F_3^{B}(0)\right],
  \label{eq:q2z}
\end{align}
which reveals how much the charge distribution of a baryon is deformed
from a spherical shape. If $\mathcal{Q}_B$ has a negative value
($\mathcal{Q}_B<0$), then the baryon takes a cushion shape, whereas if
$\mathcal{Q}_B$ is positive ($\mathcal{Q}_B>0$), then it looks like a
rugby-ball shape. 

We want to mention that the $M3$ form factors vanish in the present
work. In fact, any chiral solitonic approaches yield the null results
of the $M3$ form factors because of the hedgehog
structure~\cite{Kim:2019gka}. However, the experimental data on $M3$
is absent to date and its value should be very tiny even if it is
measured. In fact, one could compute the $M3$ form factors if one
takes into account the next-to-next-to-leading order in the $1/N_c$
expansion. This means that the $M3$ form factors 
should be strongly suppressed in the large $N_c$ limit. Thus, we will
focus in the present work on the $E2$ form factors of the baryon
sextet with spin 3/2.  

The SU(3) $\chi$QSM is constructed based on the following low-energy
effective partition function in Euclidean space, defined by 
\begin{align}
\label{eq:partftn}
\mathcal{Z}_{\chi\mathrm{QSM}} = \int \mathcal{D}\psi \mathcal{D}
  \psi^\dagger \mathcal{D} U \exp\left[-\int d^4 x \psi^\dagger D(U)
  \psi\right]  = \int \mathcal{D} U \exp (-S_{\mathrm{eff}}),  
\end{align}
where $\psi$ and $U$ denote respectively the quark and
pseudo-Nambu-Goldstone boson fields. Having integrated over quark
fields, we can express the partition function in terms of the
effective chiral action $S_{\mathrm{eff}}$, which is defined by 
\begin{align}
S_{\mathrm{eff}}(U) \;=\; -N_{c}\mathrm{Tr}\ln(i\rlap{/}{\partial} + i
  MU^{\gamma_{5}} + i   \hat{m})\,, 
\label{eq:echl}
\end{align}
where $\mathrm{Tr}$ represents the functional trace running over 
spacetime and all relevant internal spaces. The $N_c$ denotes the
number of colors. $M$ is the dynamical quark mass that arises from
spontaneous symmetry breaking of chiral symmetry. $U^{\gamma_5}$
represents the chiral field that consists of the
pseudo-Nambu-Goldstone (pNG) fields $\pi^a$, $a=1,\cdots 8$, which is
expressed as 
\begin{align}
U^{\gamma_5} = \exp(i\gamma_5\pi^a \lambda^a) = \frac{1+\gamma_5}{2} U
  + \frac{1-\gamma_5}{2} U^\dagger 
\end{align}
with 
\begin{align}
U = \exp(i\pi^a \lambda^a)   .
\end{align}
We assume isospin symmetry,
i.e., $m_{\mathrm{u}}=m_{\mathrm{d}}$. The average mass of
the up and down quarks is defined by $\overline{m}=(m_{\mathrm{u}} +
m_{\mathrm{d}})/2$. Then, the matrix of the current quark masses is
written as $\hat{m} = \mathrm{diag}(\overline{m},\, \overline{m},\,
m_{\mathrm{s}}) = \overline{m} +\delta m$. $\delta m$ is written as 
\begin{align}
\delta m  \;=\; \frac{-\overline{m} + m_{s}}{3}\bm{1} +
\frac{\overline{m} - m_{s}}{\sqrt{3}} \lambda^{8} =
m_{1} \bm{1} + m_{8} \lambda^{8}\,,
\label{eq:deltam}
\end{align}
where $m_1$ and $m_8$ denote the singlet and octet components of the  
current quark masses, defined by 
$m_1=(-\overline{m} +m_{\mathrm{s}})/3 $ and $m_8=(\overline{m}
-m_{\mathrm{s}})/\sqrt{3}$, respectively. 
The single-quark Hamiltonian $h(U)$ is defined by 
\begin{align}
h(U) \;=\;
i\gamma_{4}\gamma_{i}\partial_{i}-\gamma_{4}MU^{\gamma_{5}} -
\gamma_{4} \overline{m}\, .
\label{eq:diracham}  
\end{align}
Since the pion field has flavor indices, one has to combine a
minimal symmetric ansatz will be the hedgehog ansatz with which the
flavor indices can be coupled to three-dimensional spatial axes. 
The pion fields are then expressed in terms of a single function $P(r)$,
which is called the profile function, as follows
\begin{align}
\pi^a(\bm{x}) = n^a P(r) 
\end{align}
with $n^a = x^a/r$.  
Then the SU(2) chiral field is written as   
\begin{align}
U_{\mathrm{SU(2)}}^{\gamma_5} \;=\; \exp(i\gamma^{5}\hat{\bm{n}}\cdot
\bm{\tau} P(r))
\;=\; \frac{1+\gamma^{5}}{2}U_{\mathrm{SU(2)}} +
  \frac{1-\gamma^{5}}{2}U_{\mathrm{SU(2)}}^{\dagger}
\label{eq:embed}
\end{align}
with $U_{\mathrm{SU(2)}}=\exp(i\hat{\bm{n}}\cdot \bm{\tau} P(r))$. 
The SU(3) chiral field can be constructed by Witten's
trivial embedding~\cite{Witten:1983tx}   
\begin{align}
U^{\gamma_{5}}(x) \;=\; \left(\begin{array}{lr}
U_{\mathrm{SU(2)}}^{\gamma_{5}}(x) & 0\\
0 & 1
\end{array}\right),
\end{align}
which preserves the hedgehog ansatz. 

Integration over $U$ in Eq.~\eqref{eq:partftn} quantizes the pNG
fields. In the large $N_c$ limit, the meson mean-field approximation
is justified~\cite{Witten:1979kh,Witten:1983tx}. Thus, we can carry
out the integration over $U$ in Eq.~\eqref{eq:partftn} around the
saddle point, where $\delta S_{\mathrm{eff}}/\delta P(r) =0$ is
satisfied. This saddle-point approximation yields the equation of
motion that can be solved self-consistently. The solution provides the
self-consistent profile function $P_c(r)$ of the chiral soliton. A
detailed method of the self-consistent procedure can be found in
Ref.~\cite{Christov:1995vm}. 

While the quantum fluctuations of the self-consistent pion fields can
be ignored by the large $N_c$ argument, the fluctuations along the
direction of both the rotational and translational zero modes cannot
be ignored, since they are not at all small. Note that rotational and
translational zero modes are related to rotational and translational
symmetries. Thus, the zero modes can be taken into account by the
following rotational and translational transformations 
\begin{align}
\tilde{U}(\bm{x}, t) = A(t) U[\bm{x}- \bm{Z}(t)] A^\dagger, 
\end{align}
where $A(t)$ is an SU(3) unitary  matrix. So, the functional integral
over $U$ can be approximated by those over zero modes 
\begin{align}
\int DU [\cdots] \approx \int DA D\bm{Z}[\cdots].  
\end{align}
 The integration over translational zero modes will naturally give the
 Fourier transform of the EM densities. We refer to 
Ref.~\cite{Kim:1995mr} for a detailed description of the zero-mode 
quantization in the present scheme. 

Having carried out the zero-mode quantization, we obtain the
collective Hamiltonian as     
\begin{align}
H_{\mathrm{coll}} = H_{\mathrm{sym}} + H_{\mathrm{sb}},   
\end{align}
where
\begin{align}
  \label{eq:Hamiltonian}
H_{\mathrm{sym}} &= M_{\mathrm{cl}} + \frac1{2I_1} \sum_{i=1}^3
                   J_i^2 + \frac1{2I_2} \sum_{p=4}^7 J_p^2,\;\;\;
H_{\mathrm{sb}} = \alpha D_{88}^{(8)} + \beta \hat{Y} +
  \frac{\gamma}{\sqrt{3}} \sum_{i=1}^3 D_{8i}^{(8)} \hat{J}_i.
\end{align}
$I_1$ and $I_2$ denote the moments of inertia for the soliton, of
which the explicit expressions can be found in Appendix~\ref{app:A}. The
parameters $\alpha$, $\beta$, and $\gamma$ for heavy baryons arise
from the breaking of flavor SU(3) symmetry, which are defined by  
\begin{align}
\alpha=\left (-\frac{\overline{\Sigma}_{\pi N}}{3m_0}+\frac{
  K_{2}}{I_{2}}\overline{Y}  
\right )m_{\mathrm{s}},
 \;\;\;  \beta=-\frac{ K_{2}}{I_{2}}m_{\mathrm{s}}, 
\;\;\;  \gamma=2\left ( \frac{K_{1}}{I_{1}}-\frac{K_{2}}{I_{2}} 
 \right ) m_{\mathrm{s}}.
\label{eq:alphaetc}  
\end{align}
where $K_{1,\,2}$ are the anomalous moments of inertia, of which the
expressions are found in Appendix~\ref{app:A}. Note that the 
number of light valence quarks for a singly heavy baryon is $N_c-1$.
This means that the expression for the valence part of
$\overline{\Sigma}_{\pi N}$ contains also $N_c-1$ in place of $N_c$. 
It can be related to the $\pi N$ sigma term as follows:
$\overline{\Sigma}_{\pi N} = (N_c-1)N_c^{-1} \Sigma_{\pi N}$. The 
detailed expressions for the moments of inertia and
$\overline{\Sigma}_{\pi N}$ are given in Ref.~\cite{Kim:2018xlc}. 

The presence of the symmetry-breaking part in the collective
Hamiltonian, $H_{\mathrm{sb}}$, have the baryon wavefunctions mixed
with those in higher SU(3) representations. In the present case, the
collective wavefunctions for the baryon antitriplet 
($J=0$) and the sextet ($J=1$) are obtained respectively 
as~\cite{Kim:2018xlc}    
\begin{align}
&|B_{\overline{\bm3}_{0}}\rangle = |\overline{\bm3}_{0},B\rangle + 
p^{B}_{\overline{15}}|\overline{\bm{15}}_{0},B\rangle, \;\;\;
|B_{\bm6_{1}}\rangle = |{\bm6}_{1},B\rangle +
  q^{B}_{\overline{15}}|{\overline{\bm{15}}}_{1},B 
\rangle + q^{B}_{\overline{24}}|{
{\overline{\bm{24}}}_{1}},B\rangle,
\label{eq:mixedWF1}
\end{align}
with the mixing coefficients
\begin{eqnarray}
p_{\overline{15}}^{B}
\;\;=\;\;
p_{\overline{15}}\left[\begin{array}{c}
-\sqrt{15}/10\\
-3\sqrt{5}/20
\end{array}\right], 
& 
q_{\overline{15}}^{B}
\;\;=\;\;
q_{\overline{15}}\left[\begin{array}{c}
\sqrt{5}/5\\
\sqrt{30}/20\\
0
\end{array}\right], 
& 
q_{\overline{24}}^{B}
\;\;=\;\;
q_{\overline{24}}\left[\begin{array}{c}
-\sqrt{10}/10\\
-\sqrt{15}/10\\
-\sqrt{15}/10
\end{array}\right],
\label{eq:pqmix}
\end{eqnarray}
in the basis $\left[\Lambda_{Q},\;\Xi_{Q}\right]$ for the
antitriplet and $\left[\Sigma_{Q},\;
  \Xi_{Q}^{\prime},\;\Omega_{Q}\right]$ for the sextets with both spin
1/2 and 3/2. The parameters $p_{\overline{15}}$, $q_{\overline{15}}$, and
$q_{\overline{24}}$ are explicitly written as
\begin{eqnarray}
p_{\overline{15}}
\;\;=\;\;
\frac{3}{4\sqrt{3}}\alpha {I}_{2},
& 
q_{\overline{15}}
\;\;=\;\;
{\displaystyle -\frac{1}{\sqrt{2}}}
\left(\alpha+\frac{2}{3}\gamma\right)
{I}_{2},
& 
q_{\overline{24}}\;\;=\;\;
\frac{4}{5\sqrt{10}}
\left(\alpha-\frac{1}{3}\gamma\right) I_{2}.
\label{eq:pqmix2}
\end{eqnarray}
The collective wavefunction for the soliton with $(N_c-1)$ valence
quarks is then obtained in terms of the SU(3) Wigner $D$ functions 
\begin{align}
  \label{eq:SolitonWF1}
\psi_{(\nu;\, F),(\overline{\nu};\,\overline{S})}(R) =
  \sqrt{\mathrm{dim}(\nu)} (-1)^{Q_S} [D_{F\,S}^{(\nu)}(R)]^*,
\end{align}
where $\mathrm{dim}(\nu)$ represents the dimension of the
representation $\nu$ and $Q_S$ a charge corresponding to the soliton
state $S$, i.e., $Q_S=J_3+Y'/2$. $F$ and $S$ stand for the flavor and
spin quantum numbers corresponding to the soliton for the singly heavy
baryon. Finally, the complete wavefunction for a singly heavy baryon
can be derived by coupling the soliton wavefunction to the heavy quark
spinor  
\begin{align}
\Psi_{B_{Q}}^{(\mathcal{R})}(R)
 =  \sum_{J_3,\,J_{Q3}} 
C_{\,J,J_3\, J_{Q}\,J_{Q3}}^{J'\,J_{3}'}
\;\mathbf{\chi}_{J_{Q3}}
\;\psi_{(\nu;\,Y,\,T,\,T_{3})(\overline{\nu};\,Y^{\prime},\,J,\,J_3)}(R),
\label{eq:HeavyWF}
\end{align}
where $\chi_{J_{Q3}}$ denote the Pauli spinors for the heavy quark and
$C_{\,J,J_3\, J_{Q}\,J_{Q3}}^{J'\,J_{3}'}$ the Clebsch-Gordan
coefficients. 

The matrix elements of the EM current~\eqref{eq:MatrixEl1} can be
computed within the $\chi$QSM by representing them in terms of the
functional integral in Euclidean space,
\begin{align}
\langle B,\,p'| J_\mu(0) |B,\,p\rangle  &= \frac1{\mathcal{Z}}
  \lim_{T\to\infty} \exp\left(i p_4\frac{T}{2} - i p_4'
  \frac{T}{2}\right) \int d^3x d^3y \exp(-i \bm{p}'\cdot \bm{y} + i
  \bm{p}\cdot \bm{x}) \cr
& \hspace{-1cm} \times \int \mathcal{D}U\int \mathcal{D} \psi \int
  \mathcal{D} \psi^\dagger J_{B}(\bm{y},\,T/2) \psi^\dagger(0)
  \gamma_4\gamma_\mu \hat{Q} \psi(0) J_B^\dagger (\bm{x},\,-T/2)
  \exp\left[-\int d^4 z   \psi^\dagger iD(U) \psi\right],  
\label{eq:correlftn}
\end{align}
where the baryon states $|B,\,p\rangle$ and $\langle B,\,p'|$ are, 
respectively, defined by 
\begin{align}
|B,\,p\rangle &= \lim_{x_4\to-\infty}   \exp(i p_4 x_4)
                \frac1{\sqrt{\mathcal{Z}}} \int d^3 x
                \exp(i\bm{p}\cdot \bm{x}) J_B^\dagger
                (\bm{x},\,x_4)|0\rangle,\cr
\langle B,\,p'| &= \lim_{y_4\to\infty}   \exp(-i p_4' y_4)
                \frac1{\sqrt{\mathcal{Z}}} \int d^3 y
                \exp(-i\bm{p}'\cdot \bm{y}) \langle 0| J_B^\dagger 
                (\bm{y},\,y_4).
\end{align}
The heavy baryon current $J_B$ can be constructed from the $N_c-1$
valence quarks
\begin{align}
J_B(x) = \frac1{(N_c-1)!} \epsilon_{i_1\cdots i_{N_c-1}} \Gamma_{JJ_3
  TT_3 Y}^{\alpha_1\cdots \alpha_{N_c-1}} \psi_{\alpha_1 i_1} (x)
  \cdots \psi_{\alpha_{N_c-1} i_{N_c-1}}(x),  
\end{align}
where $\alpha_1\cdots \alpha_{N_c-1}$ represent spin-flavor indices
and $i_1\cdots i_{N_c-1}$ color indices. The matrices $\Gamma_{JJ_3
  TT_3 Y}^{\alpha_1\cdots \alpha_{N_c-1}}$ are taken to consider the
quantum numbers $JJ_3TT_3Y$ of the $N_c-1$ soliton. The creation
operator $J_B^\dagger$ can be constructed in a similar
way. The calculation of the baryonic correlation function given in 
Eq.~\eqref{eq:correlftn} is a tedious one, so we will present here
only the final expressions for the $E2$ form factor, As for the
detailed formalism, we refer to
Refs.~\cite{Christov:1995vm,Kim:1995mr}.  

The final expressions for the electric quadrupole form factors of the 
baryon sextet with spin 3/2 can be written as 
\begin{align}
G^{B_6}_{E{2}}(Q^{2}) &=  6 \sqrt{{5} } \frac{M^{2}_{B}}{|\bm{q}|^{2}}
 \int d^{3} z \,  j_{2}(|\bm{q}||\bm{z}|)
 {\mathcal{G}}^{B}_{E2}(\bm{z}) ,    
\label{eq:magfinal}
\end{align}
where $j_2(|\bm{q}||\bm{z}|)$ stands for the spherical Bessel function
with order 2 and the corresponding density of the $E2$ form factors is given
as 
\begin{align}
{\mathcal{G}}^{B}_{E2}(\bm{z}) =& -2 \left( \frac{3}{I_{1}} \langle
 D^{(8)}_{Q3} J_{3} \rangle_{B}  - \frac{1}{I_{1}} \langle
 D^{(8)}_{Q i} J_{i} \rangle_{B} \right)
 {\cal {I}}_{1E2} (\bm{z})  \cr    
 & + 4 m_{8}\left( \frac{K_{1}}{I_{1}} \mathcal{I}_{1E2}(\bm{z}) -
    \mathcal{K}_{1E2}(\bm{z})\right) \left( 3\langle D^{(8)}_{8 3}
    D^{(8)}_{Q 3}\rangle_{B} -\langle D^{(8)}_{8 i} D^{(8)}_{Q i}
    \rangle_{B} \right). 
\label{eq:magden}
\end{align}
The densities of $E2$ form factors
$\mathcal{I}_{1E2}$ and $\mathcal{K}_{1E2}$ can be found in
Appendix~\ref{app:A}. In the limit of $m_Q\to \infty$, the charge
distribution of the heavy quark becomes a point-like static
charge given as $\rho_Q(\bm{r})=e_Q \delta^{(3)}(\bm{r})$. This leads
to $\mathcal{Q}_{ij} = \int d^{3} r \rho_Q(\bm{r})( 3 r_{i}r_{j} -
r^{2}\delta_{ij})=0$. This implies that the $E2$ form factors of the
singly heavy baryons are solely governed by the light quarks in the
$m_Q\to \infty$ limit.  

Having calculated the matrix elements of the collective
operators in Eq.~\eqref{eq:magden}, we arrive at the final expressions
for the $E2$ form factors of the baryon sextet with spin 3/2 
\begin{align}
\mathcal{G}^{B}_{E2}(\bm{z}) &= \mathcal{G}^{B(0)}_{E2}(\bm{z})
 + \mathcal{G}^{B(\text{op})}_{E2}(\bm{z})
 + \mathcal{G}^{B(\text{wf})}_{E2}(\bm{z}),   
\label{eq:E2final}
\end{align}
where $\mathcal{G}_{E2}^{B(0)}$, $\mathcal{G}_{E2}^{B(\mathrm{op})}$,
and $\mathcal{G}_{E2}^{B(\mathrm{wf})}$ denote respectively the
symmetric terms, the flavor SU(3) symmetry-breaking ones from the
effective chiral action, and those from the mixed collective
wavefunctions, expressed explicitly as
\begin{align}
{\cal{G}}^{B_{6}(0)}_{E2}(\bm{z})& = \frac{3}{10}\frac{1}{I_{1}}
  Q_{B} \mathcal {I}_{1E2} (\bm{z}) ,
\label{eq:leading_order}
\\ 
  {\cal{G}}^{B_{6}(\text{op})}_{E2}(\bm{z}) &= -\frac{1}{405} m_{s}
  \left(\frac{K_1}{I_{1}}\mathcal{I}_{E2}(\bm{z})
-\mathcal{K}_{E2}(\bm{z})\right)
  \left(\begin{array}{c c c} 6Q_{\Sigma_{c}^{*}}+1 \\
 -24Q_{\Xi_{c}^{*}}-13 \\ 9 \end{array} \right),
\label{eq:ms_op}
\\
{\cal{G}}^{B_{6}(\text{wf})}_{E2}(\bm{z}) & = -\frac{2}{I_{1}} \left[ 
q_{\overline{15}} \left( \begin{array}{c c c}
 {-\frac{2}{9\sqrt{5}}}(3Q_{\Sigma^{*}_{c}}-4) \\
 {-\frac{1}{18\sqrt{5}}}(15Q_{\Xi^{*}_{c}}-2)\\ 
0  \end{array} \right)+
  q_{\overline{24}}\left( \begin{array}{c c c}
 {-\frac{1}{180}}(3Q_{\Sigma^{*}_{c}}+5) \\
 {-\frac{1}{90}}(3Q_{\Xi^{*}_{c}}+5) \\
 {\frac{3}{40}}Q_{\Omega^{*}_{c}}   
\end{array}
 \right) \right] \mathcal{I}_{1E2} (\bm{z}), 
\label{eq:ms_wf}
\end{align}
where $Q_{B}$ stands for the charge of the light-quark 
components of the corresponding baryons. We can derive similar sum
rules for the electric quadrupole moments of singly heavy baryons with
spin 3/2 as follows~\cite{Kim:2019gka}
\begin{align}
&\sum_{B\in\mathrm{sextet}} \mathcal{Q}_{B} = 0, \cr
&\mathcal{Q}_{\Sigma^{*0}_{c}}= \mathcal{Q}_{\Xi^{*0}_{c}} =
  \mathcal{Q}_{\Omega^{*0}_{c}} = -2 
  \mathcal{Q}_{\Sigma^{*+}_{c}}=-2 \mathcal{Q}_{\Xi^{*+}_{c}}=-\frac{1}{2}
  \mathcal{Q}_{\Sigma^{*++}_{c}}. 
\end{align}
Even though the flavor SU(3) symmetry is broken, we still can find the
following sum rules 
\begin{align}
\mathcal{Q}_{\Sigma^{*++}_{c}}-\mathcal{Q}_{\Sigma^{*+}_{c}} &=
 \mathcal{Q}_{\Sigma^{*+}_{c}}  - \mathcal{Q}_{\Sigma^{*0}_{c}}, \cr 
\mathcal{Q}_{\Sigma^{*0}_{c}}-\mathcal{Q}_{\Xi^{*0}_{c}} &=
 \mathcal{Q}_{\Xi^{*0}_{c}}  -  \mathcal{Q}_{\Omega^{*0}_{c}}, \cr  
2(\mathcal{Q}_{\Sigma^{*+}_{c}}-\mathcal{Q}_{\Xi^{*0}_{c}}) &= 
\mathcal{Q}_{\Sigma^{*++}_{c}} - \mathcal{Q}_{\Omega^{*0}_{c}}.  
\end{align}

\section{Results and discussion}
In the $\chi$QSM, there are several parameters to fix. Since the
sea-quark or Dirac-sea contributions contain divergent integrals, one
has to introduce a regularization to tame the 
divergences. In the present work, we introduce the proper-time
regularizations with the cutoff mass. This can be fixed by
using the pion decay constant $f_\pi=93$ MeV. The average mass of the
up and down current quarks $\overline{m}$ is determined by the
physical pion mass $m_\pi=140$ MeV (see Appendix~\ref{app:B} for
details). While the mass of the strange current quark $m_{\mathrm{s}}$
can be also fixed by reproducing the kaon mass, which gives
$m_{\mathrm{s}}=150$ MeV, we preferably use 
$m_{\mathrm{s}}=180$ MeV, since this value of $m_{\mathrm{s}}$ yields
the best results for the hyperon mass
splittings~\cite{Blotz:1992pw,Christov:1995vm}. The remaining
parameter is the dynamical quark mass $M$, which is the 
only free parameter of the model. However, $M=420$ MeV is known to be
the best value in reproducing various observables in the light baryon
sector~\cite{Christov:1995vm}. Thus, we will use this value also in
the present calculation.

\begin{figure}[htp]
\centering
\includegraphics[scale=0.26]{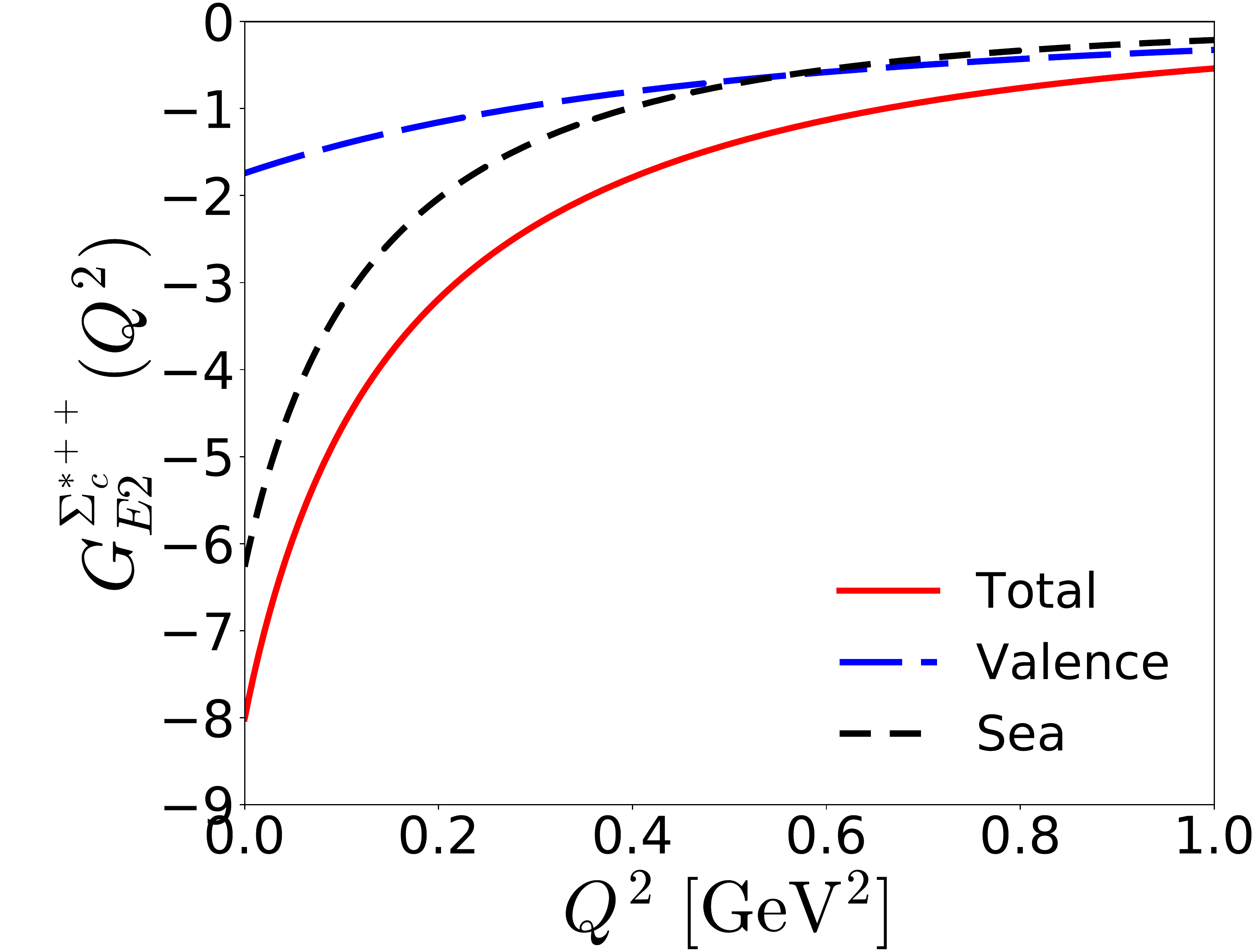}
\includegraphics[scale=0.26]{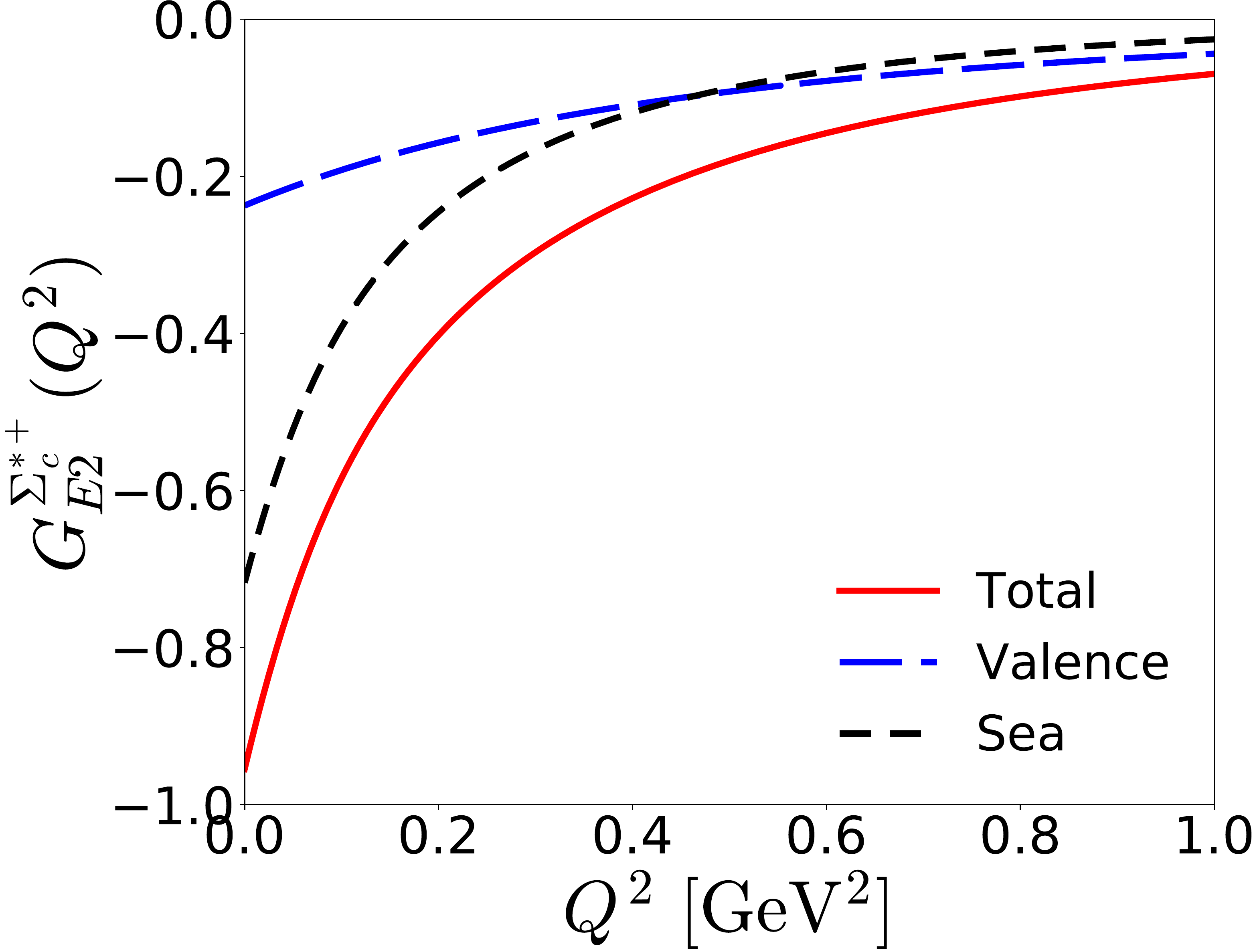}
\includegraphics[scale=0.26]{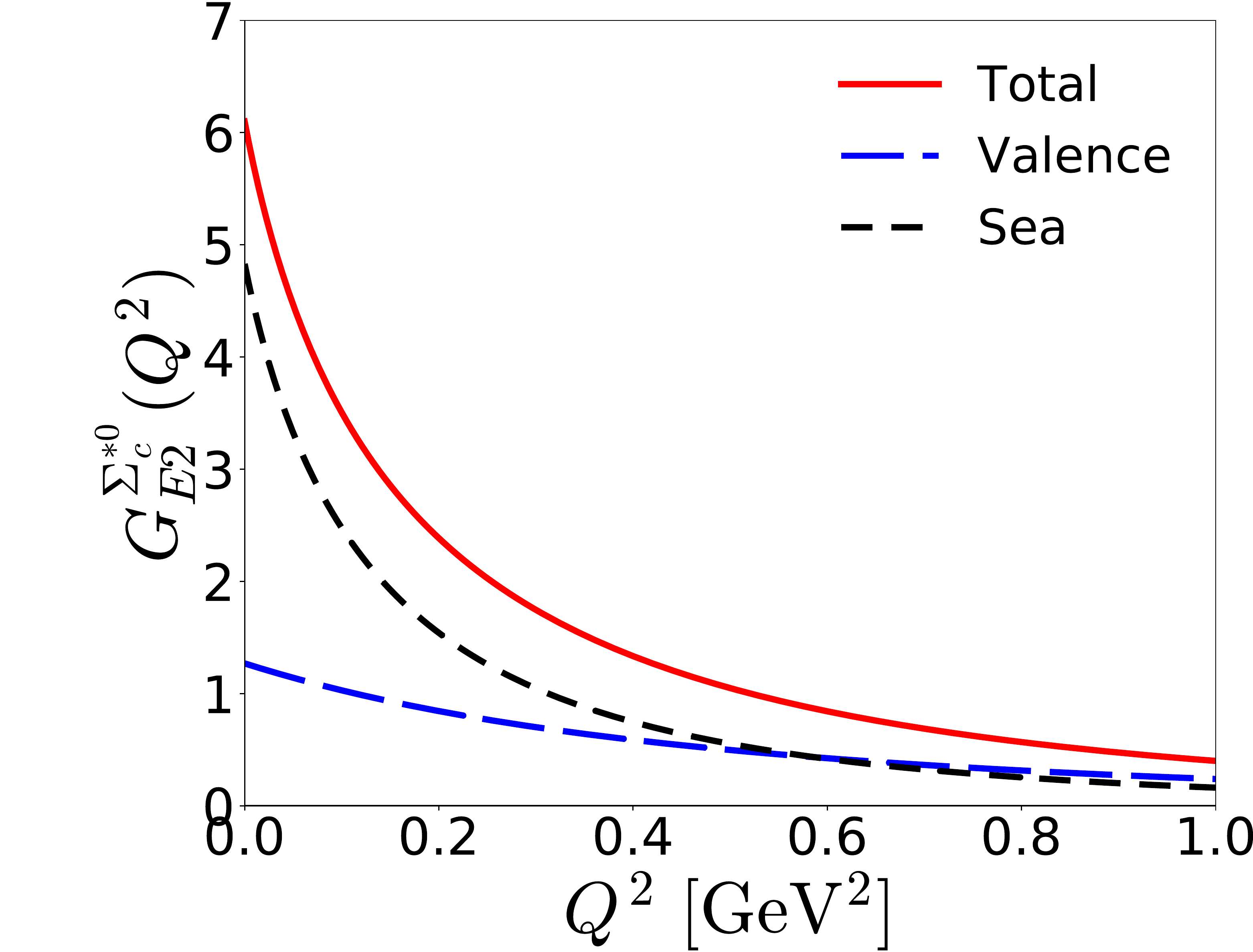}
\includegraphics[scale=0.26]{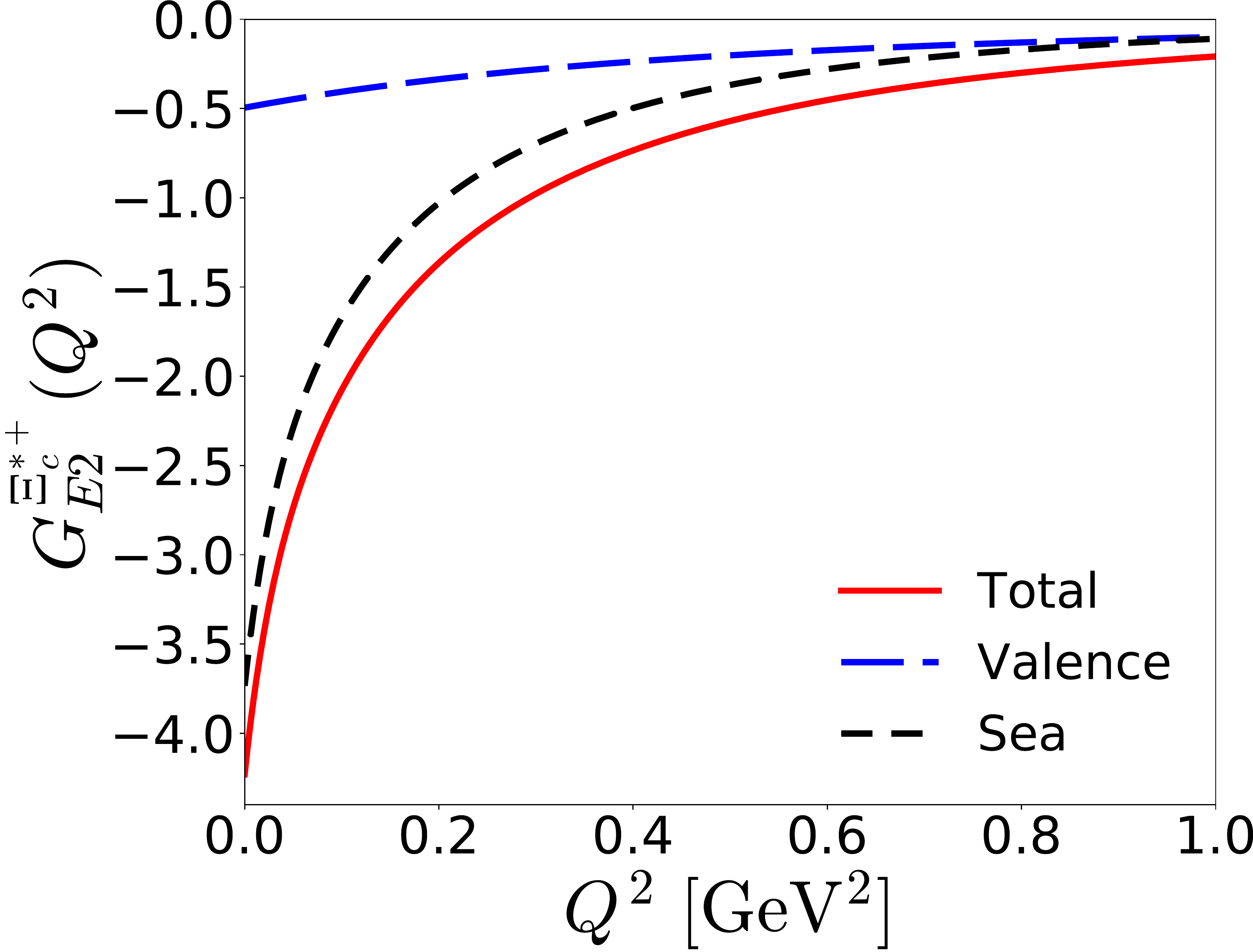}
\includegraphics[scale=0.26]{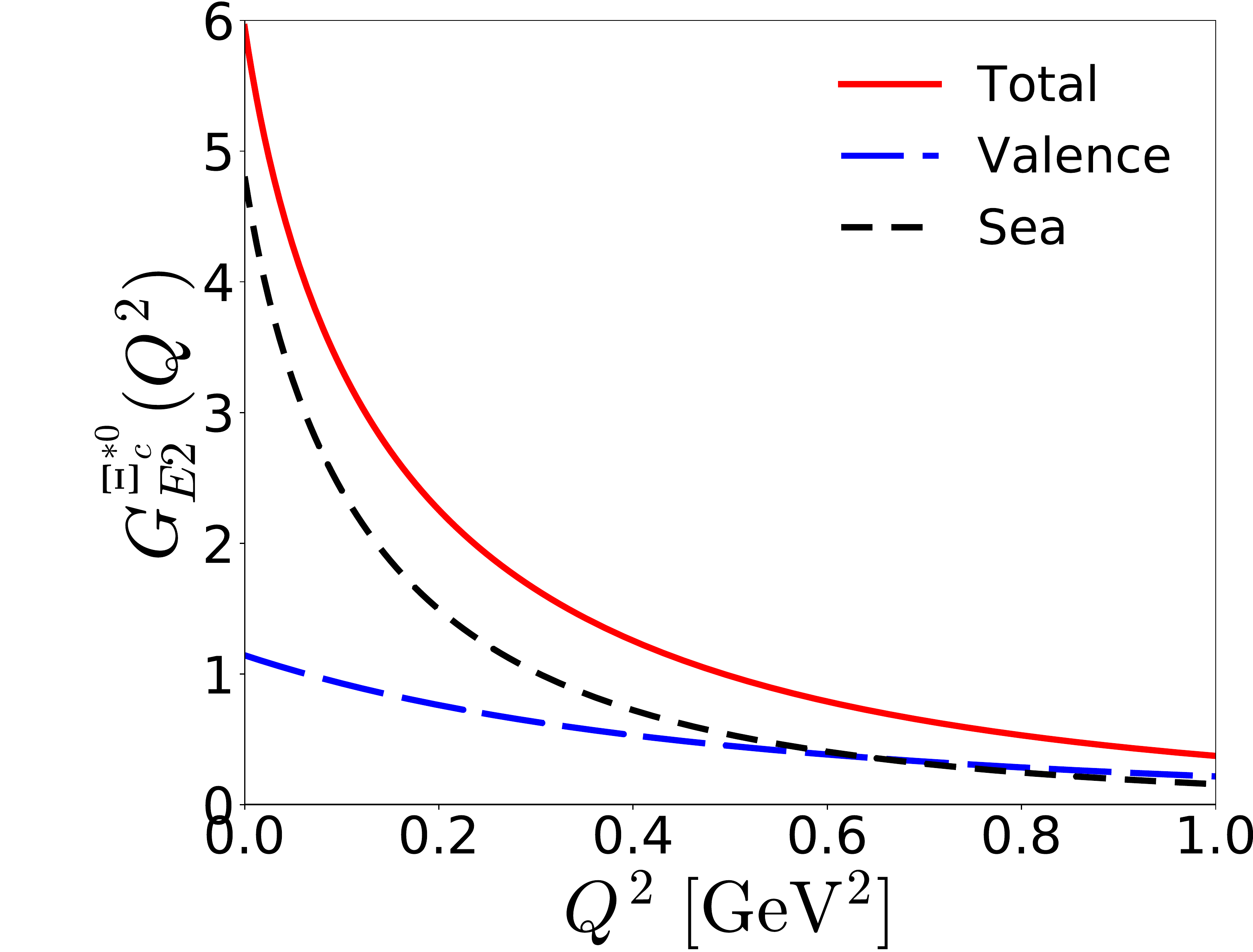}
\includegraphics[scale=0.26]{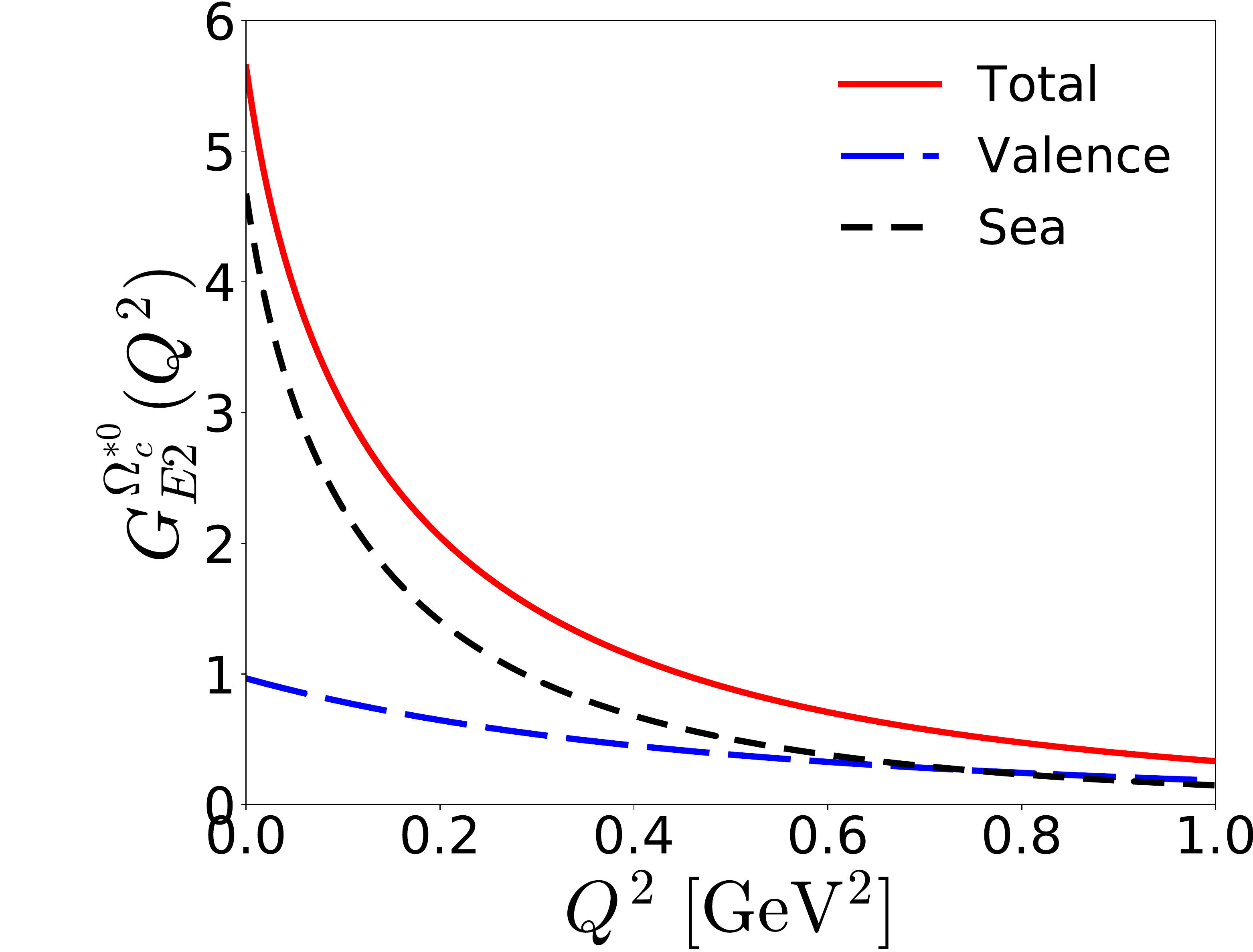}
\caption{Valence- and sea-quark contributions to the electric
  quadrupole form factors of the baryon sextet with spin 3/2. The
  long-dashed curves draw the valence-quark contributions to the $E2$
  form factors, whereas the short-dashed ones depict the sea-quark
  contributions. The solid ones represent the total results for the
  $E2$ form factors.}  
\label{fig:1}
\end{figure}
It was shown that in the calculation of the $E2$ form factors of the
baryon decuplet the sea-quark contributions turn out to be rather
important, we will first examine the valence- and sea-quark
contributions separately. In Fig.~\ref{fig:1}, we draw the numerical
results for the $E2$ form factors of the baryon sextet with spin 3/2.
As expected, the general behaviors of the valence- and sea-quark
contributions to the $E2$ form factors of the heavy singly baryons are
rather similar to those of the baryon decuplet. As shown in
Fig.~\ref{fig:1}, the valence-quark contributions decrease off mildly as
$Q^2$ increases, whereas the sea-quark or Dirac-sea contributions fall
off drastically in the smaller $Q^2$ region, so that they govern the
$Q^2$ dependence of the $E2$ form factors. In particular, the
magnitudes of the sea-quark contributions are quite larger than in the
region of smaller $Q^2$. Thus, they are the main contributions to the
electric quadrupole moments of the baryon sextet with spin
3/2. Considering the fact that the electric quadrupole moment exhibits 
how the corresponding baryon is deformed, the present results provide
certain physical implications. Recent investigations on the
gravitational form factors of baryons within the $\chi$QSM indicate
that the valence quarks are mainly located in the inner part of a
baryon, while the sea quarks lie in its outer part~\cite{Goeke:2007fp,
  Kim:2020nug}. Thus, the sea-quark contributions, which can be also
interpreted as pion clouds, mainly describe how a singly heavy baryon
with spin 3/2 is deformed. The present results are in line with what
was discussed in Ref.~\cite{Pascalutsa:2006up}, where the significance
of the pion clouds in the electric quadrupole moment of the $\Delta$
isobar was discussed . 

\begin{figure}[htp]
\centering
\includegraphics[scale=0.26]{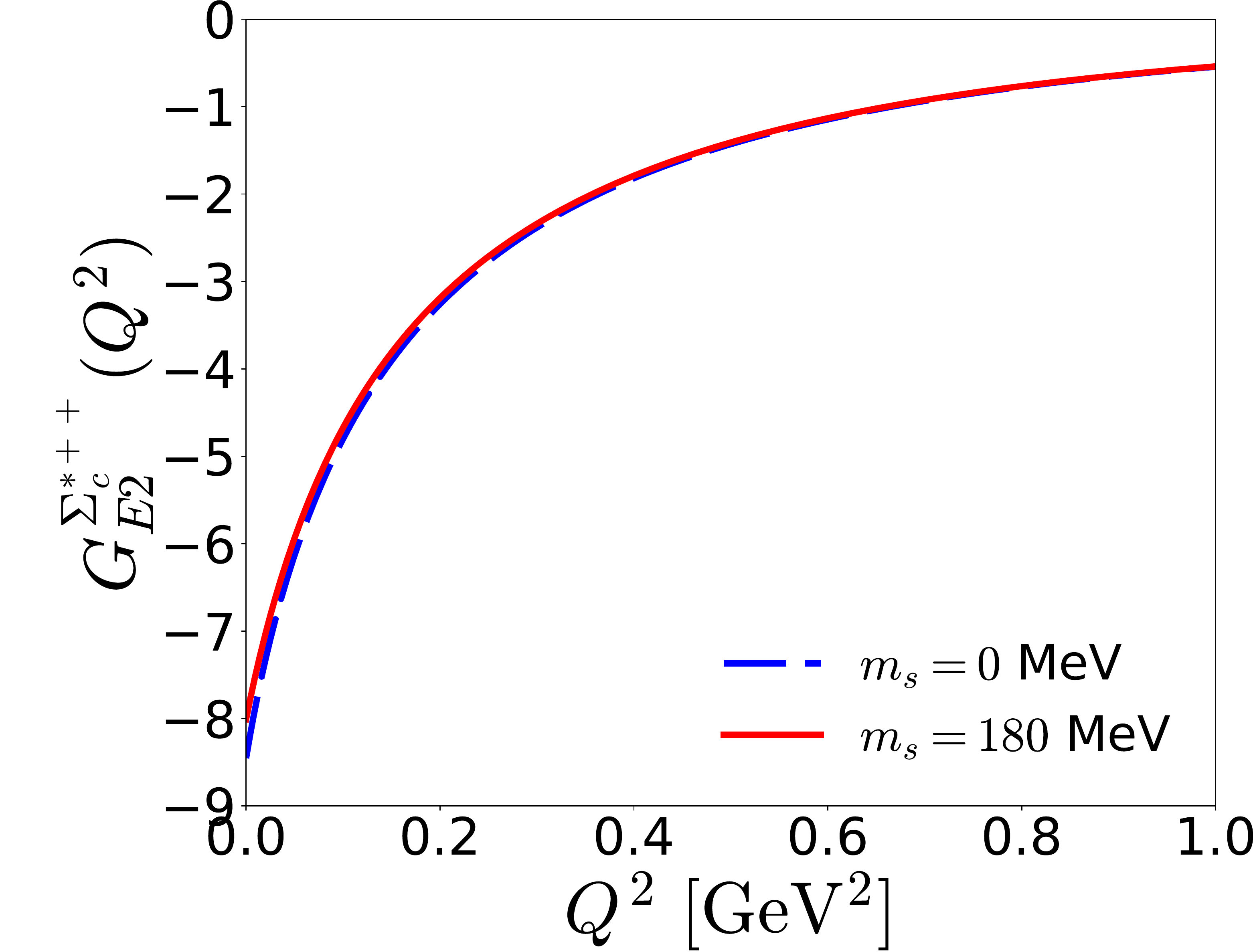}
\includegraphics[scale=0.26]{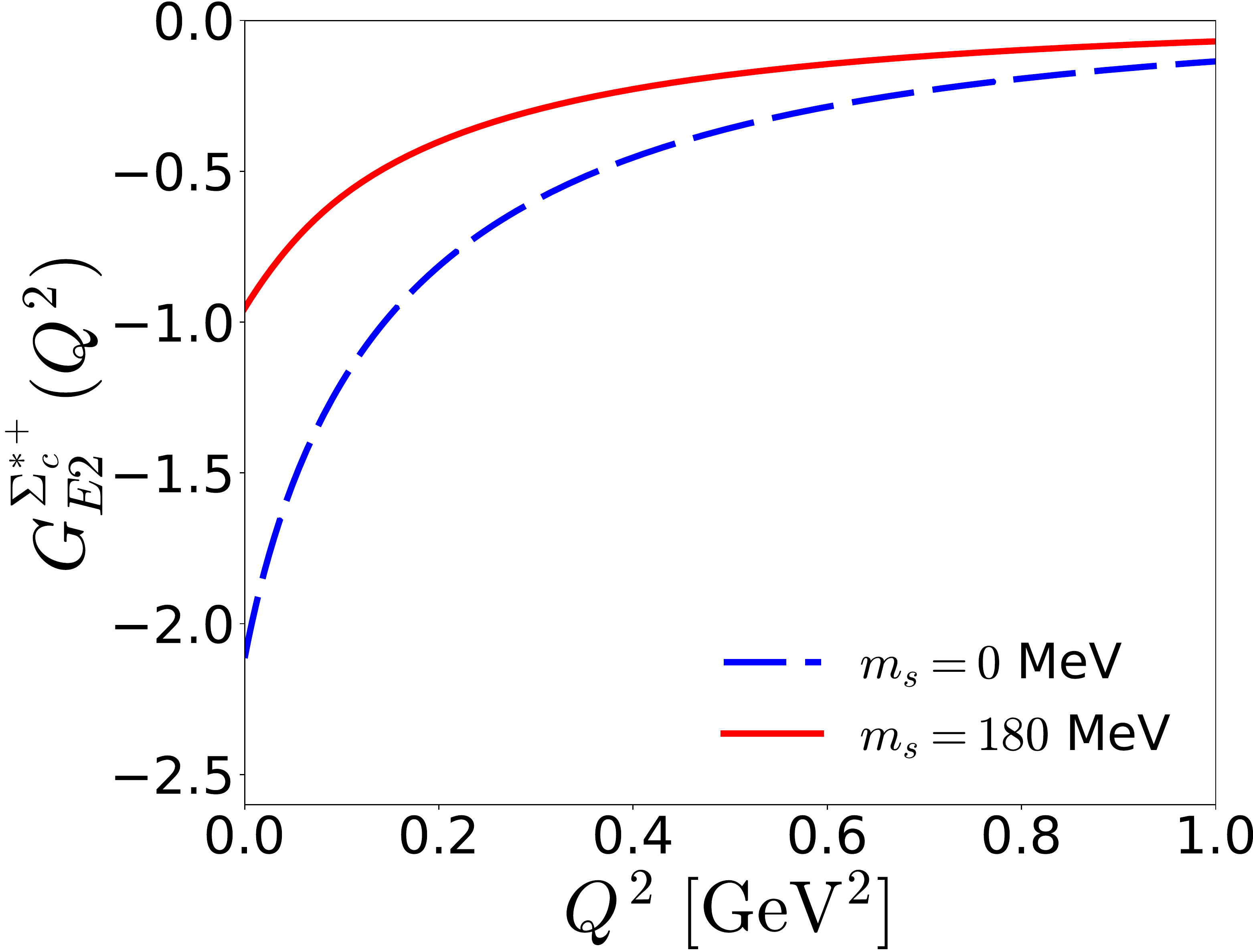}
\includegraphics[scale=0.26]{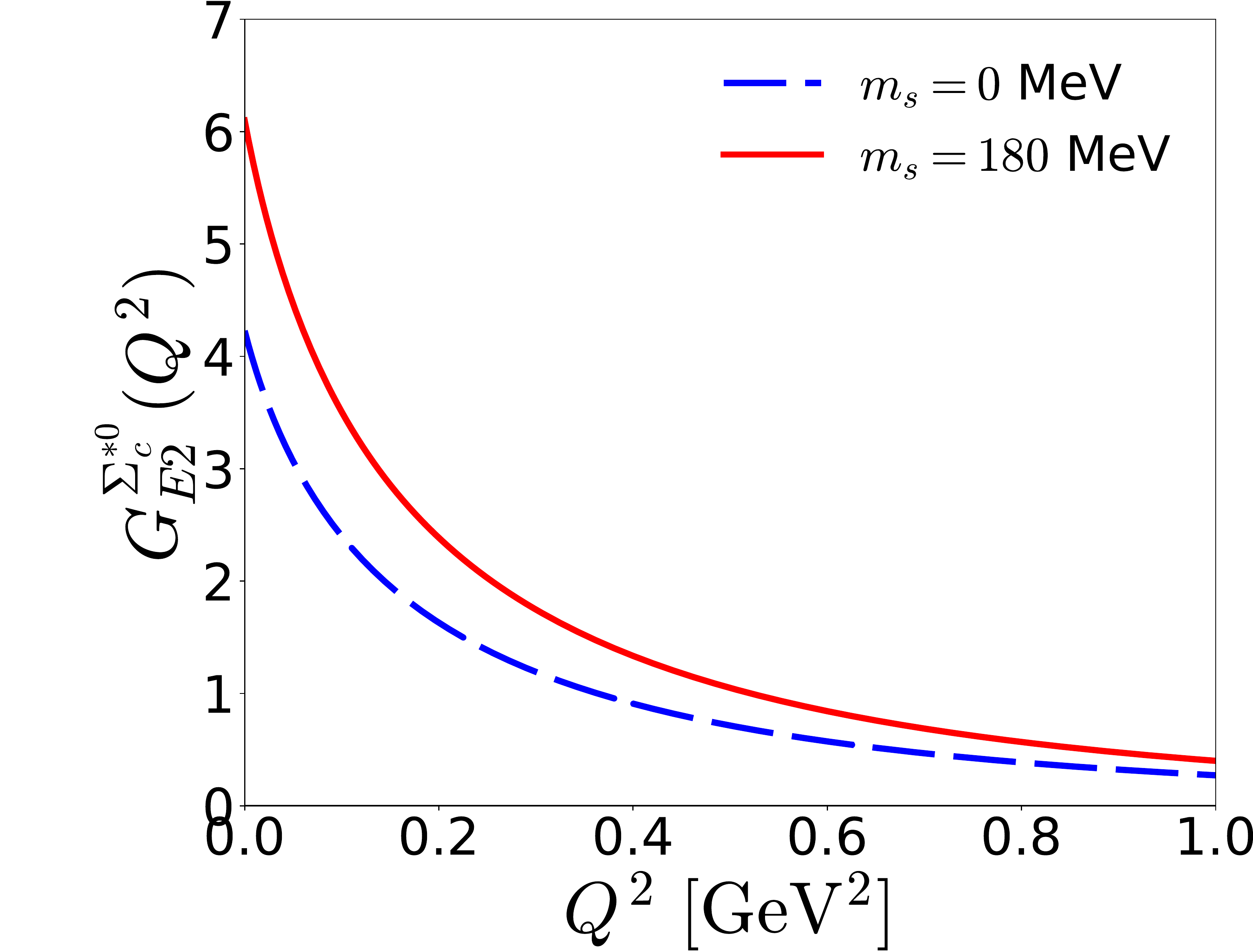}
\includegraphics[scale=0.26]{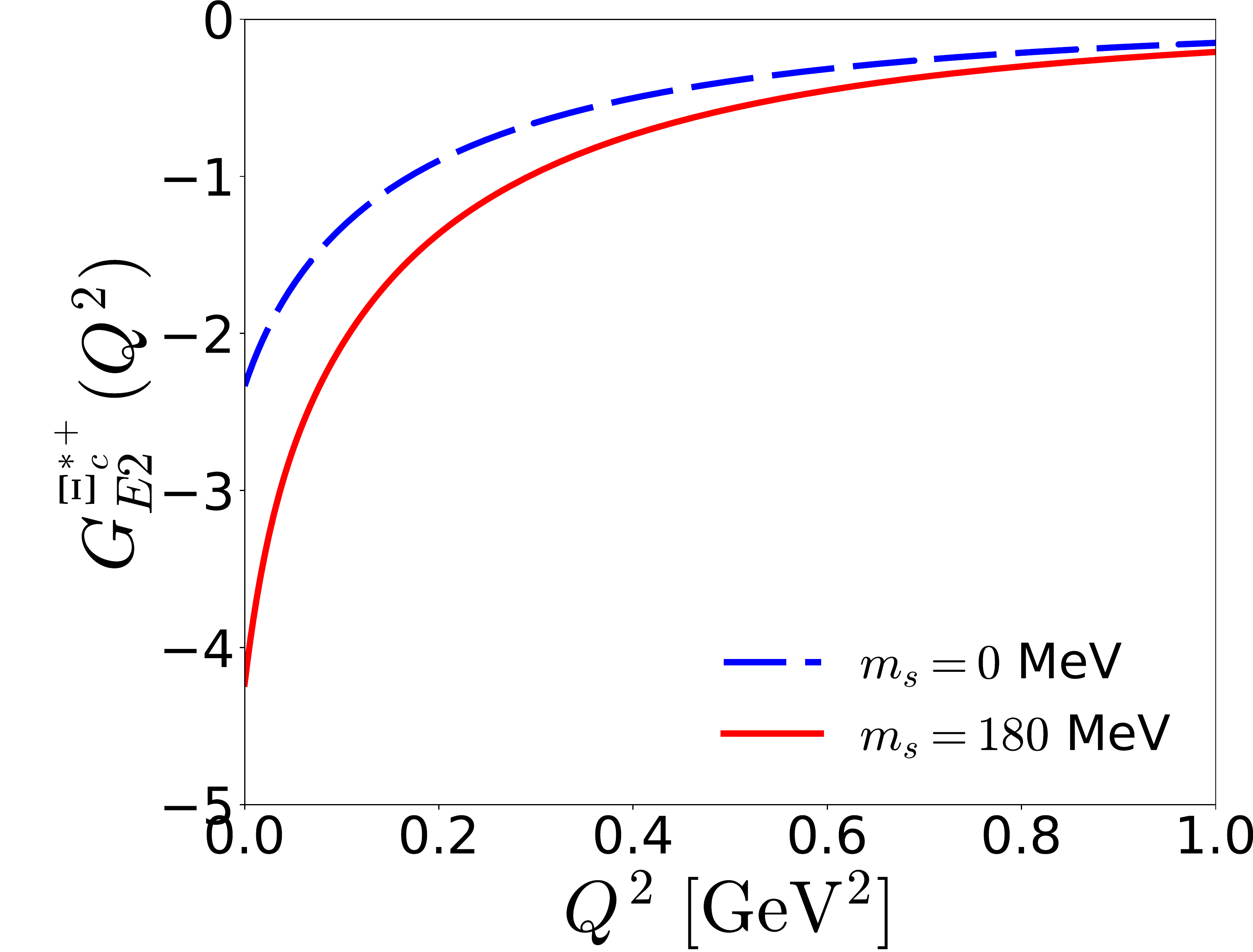}
\includegraphics[scale=0.26]{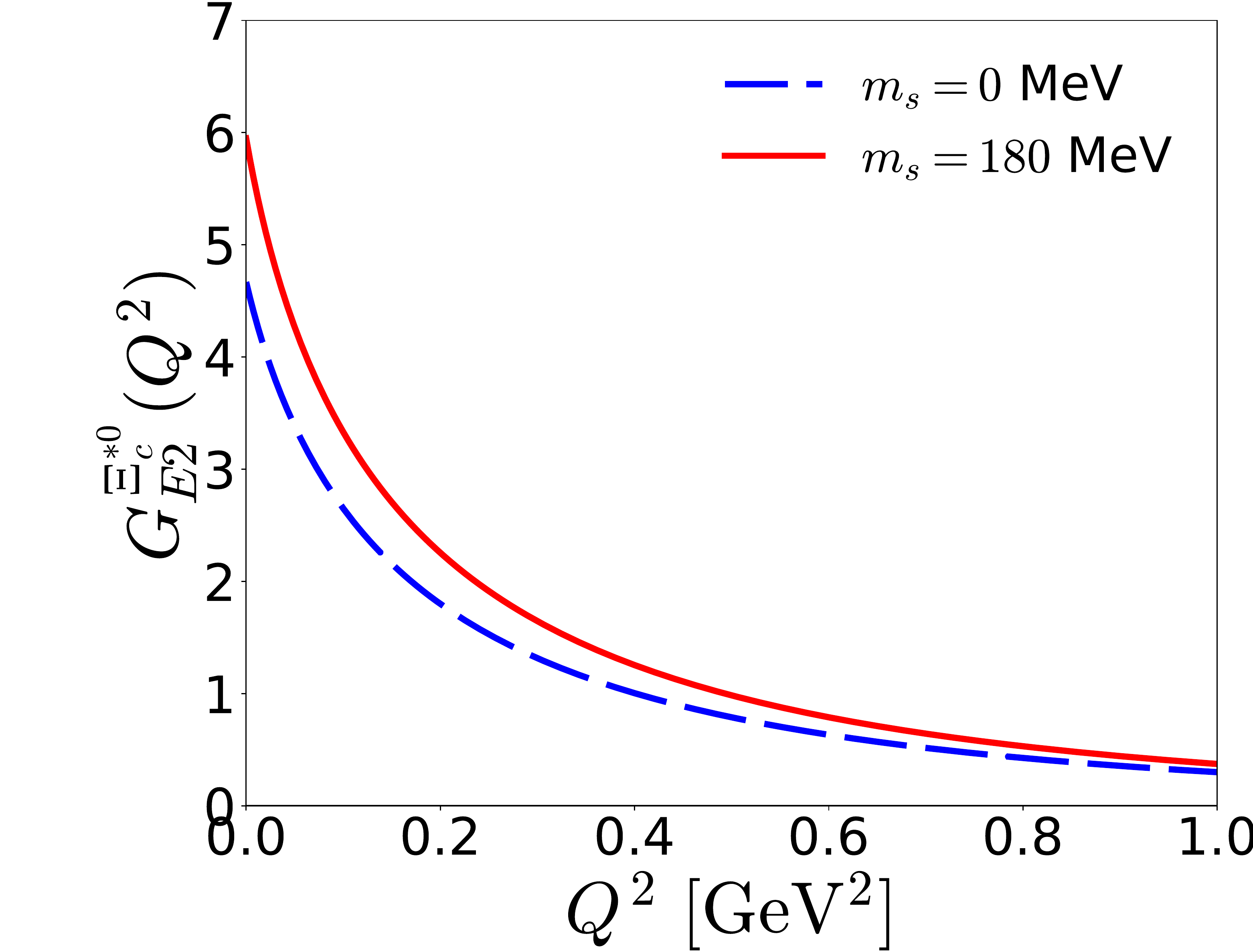}
\includegraphics[scale=0.26]{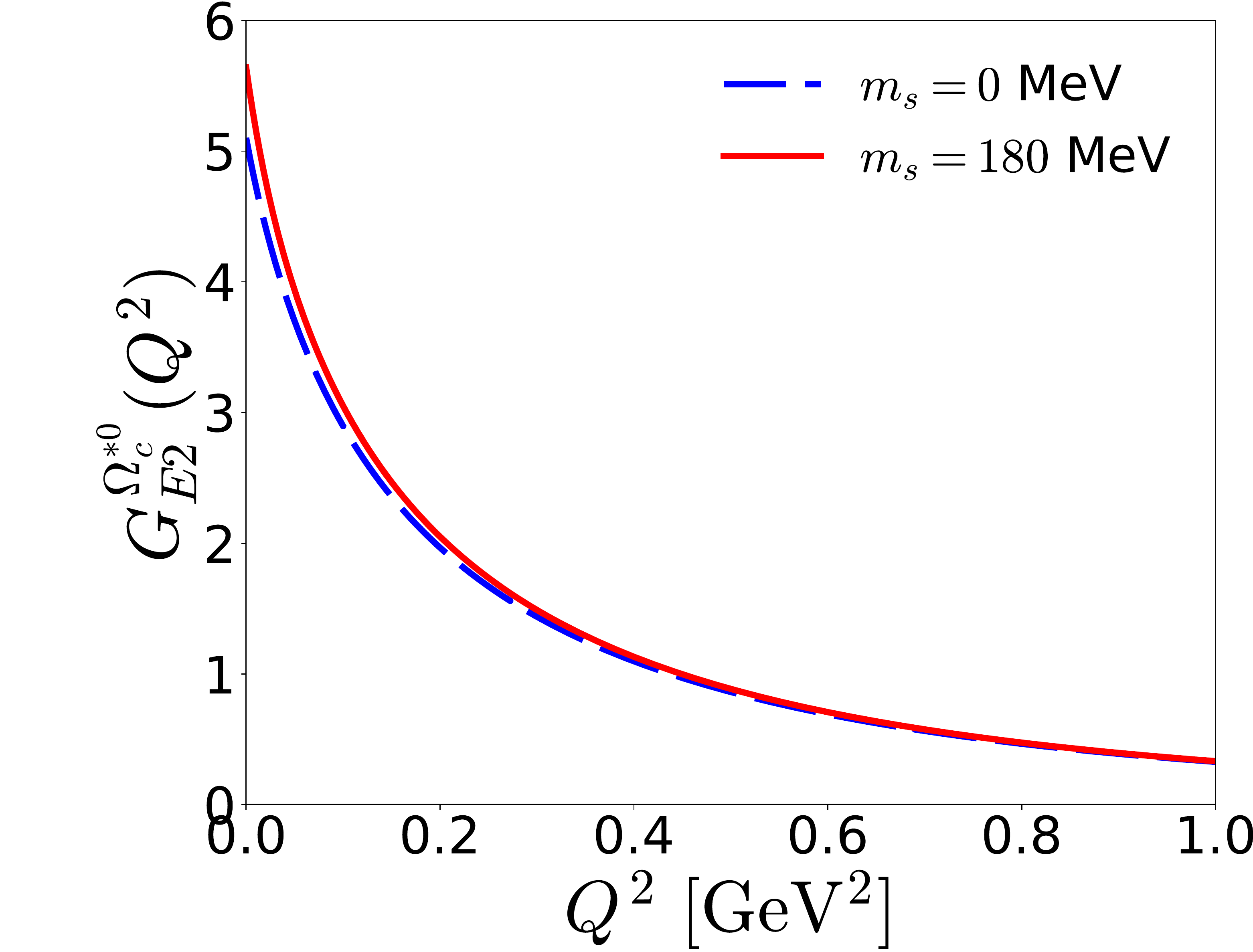}
\caption{The effects of flavor SU(3) symmetry breaking on the electric
  quadrupole form factors of the baryon sextet with spin 3/2. The
  dashed curves draw the results for the $E2$ form factors without the
$m_{\mathrm{s}}$ corrections, whereas the solid curves depict the
results with the effects of flavor SU(3) symmetry breaking taken into
account.}  
\label{fig:2}
\end{figure}
\begin{figure}[htp]
\centering
\includegraphics[scale=0.26]{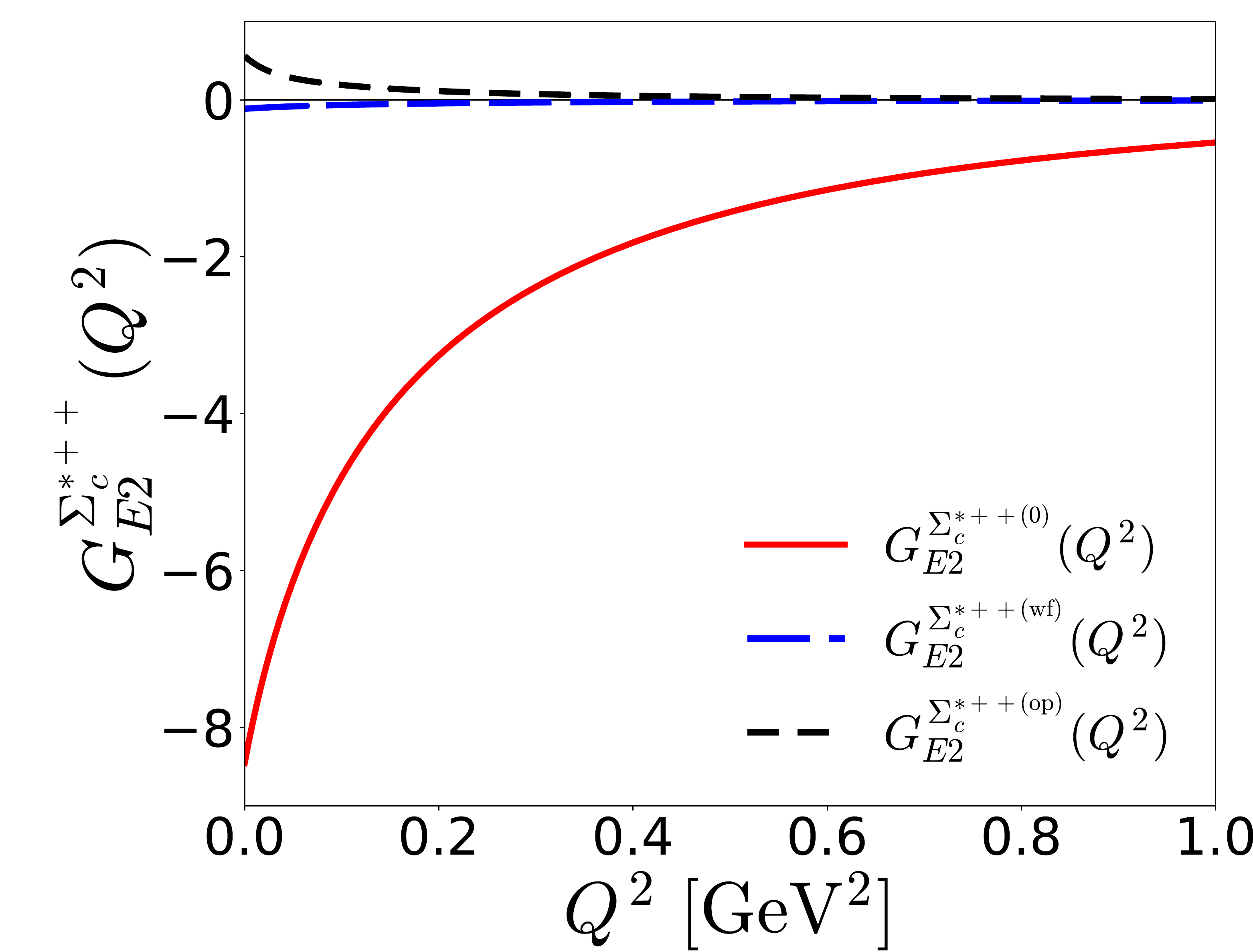}
\includegraphics[scale=0.26]{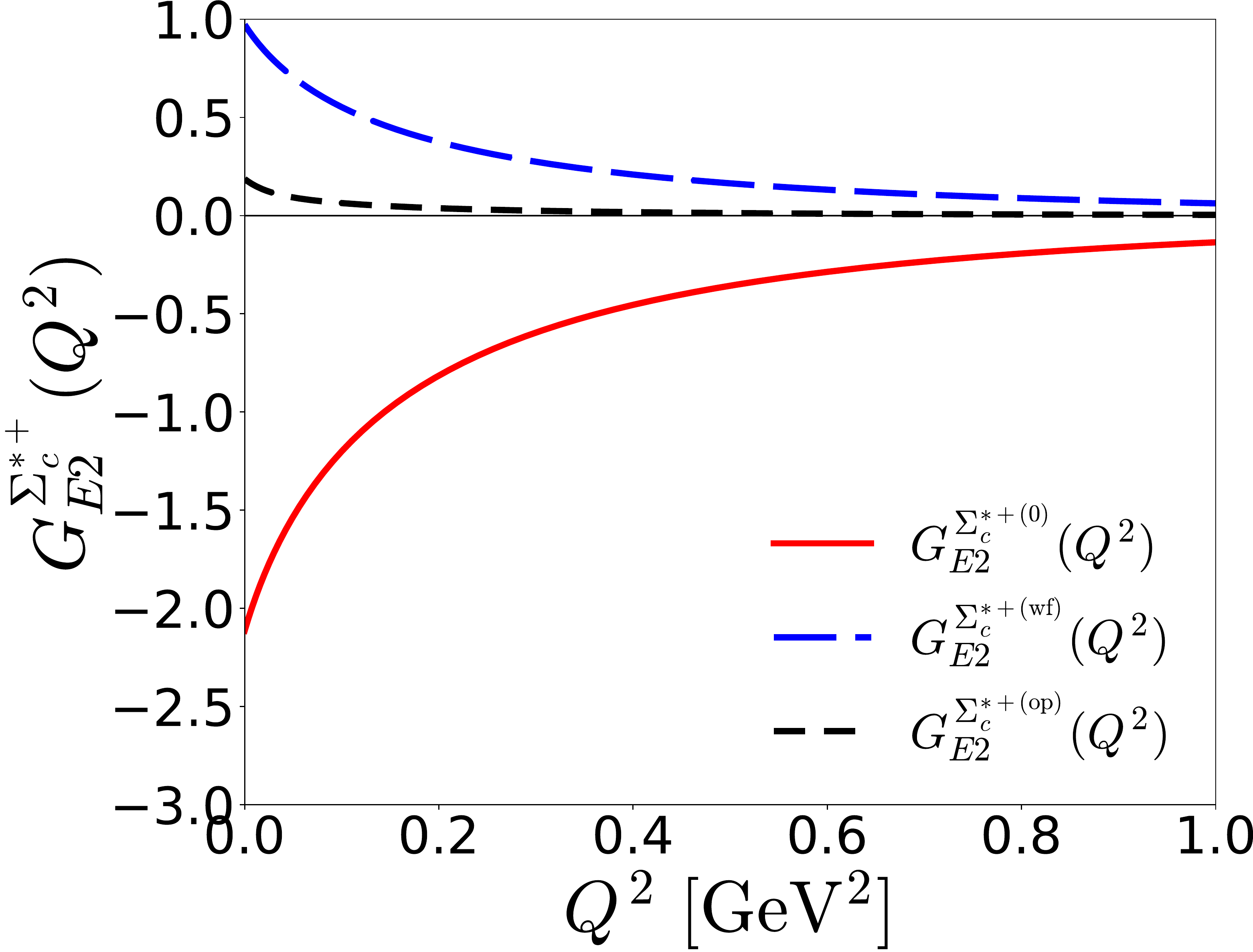}
\includegraphics[scale=0.26]{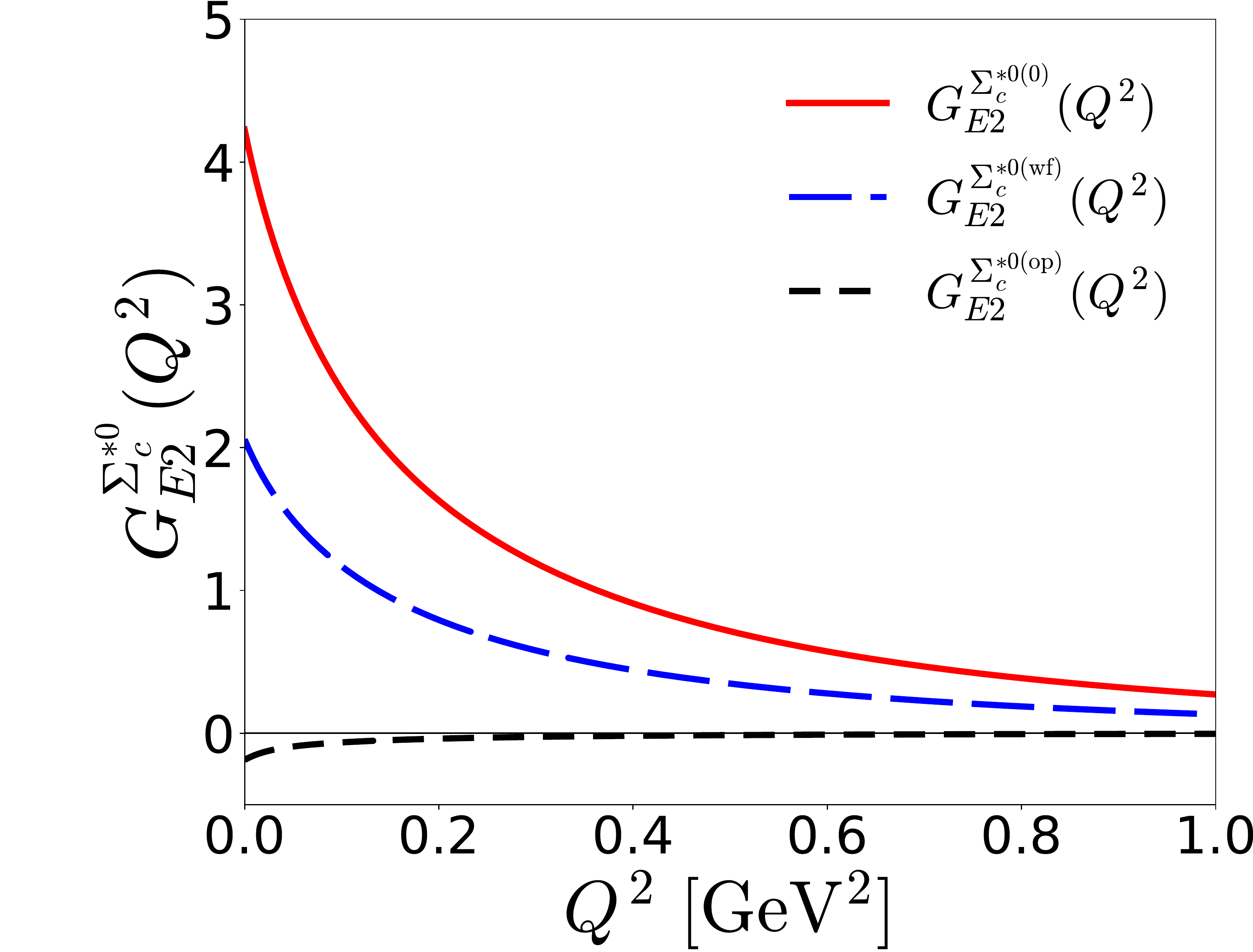}
\includegraphics[scale=0.26]{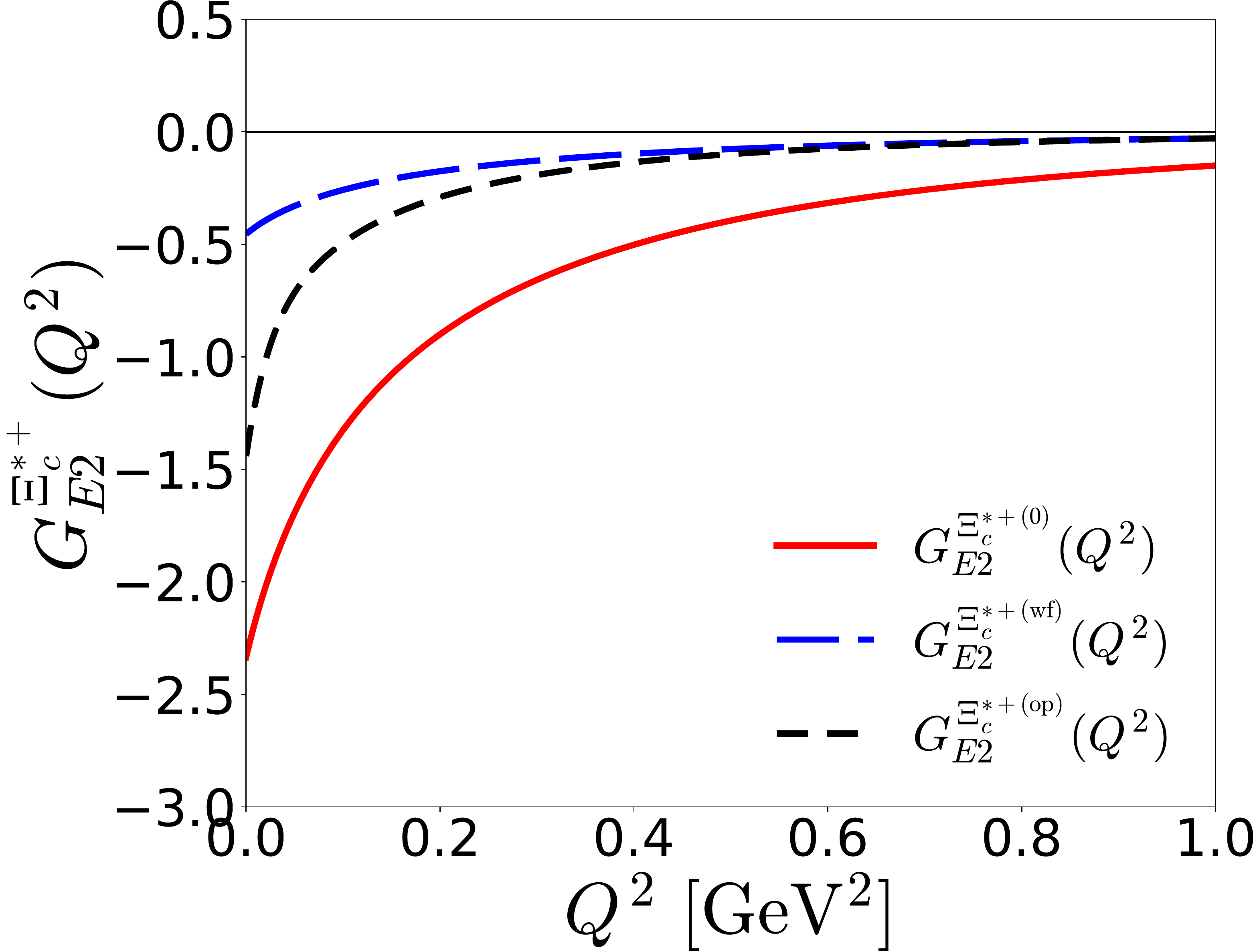}
\includegraphics[scale=0.26]{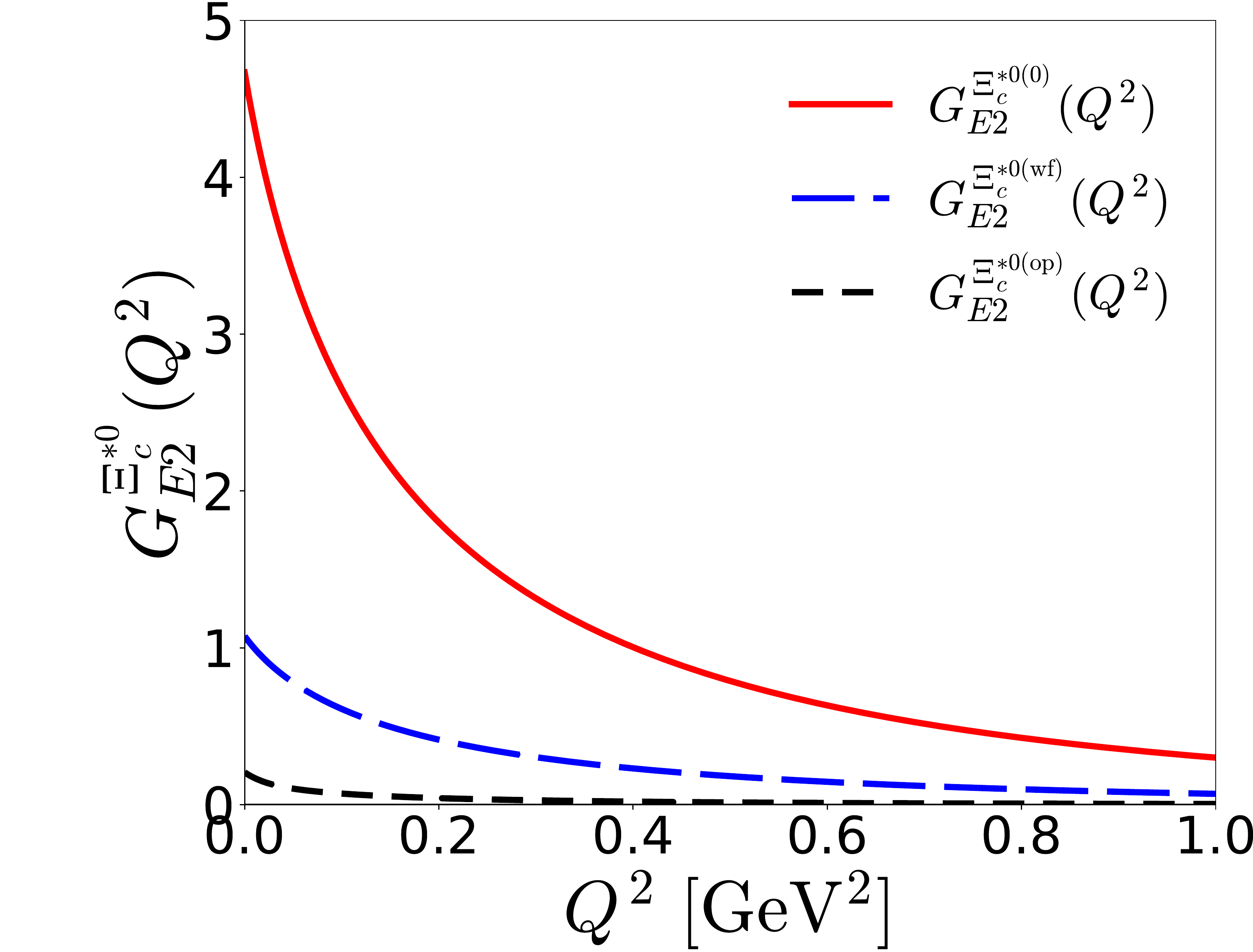}
\includegraphics[scale=0.26]{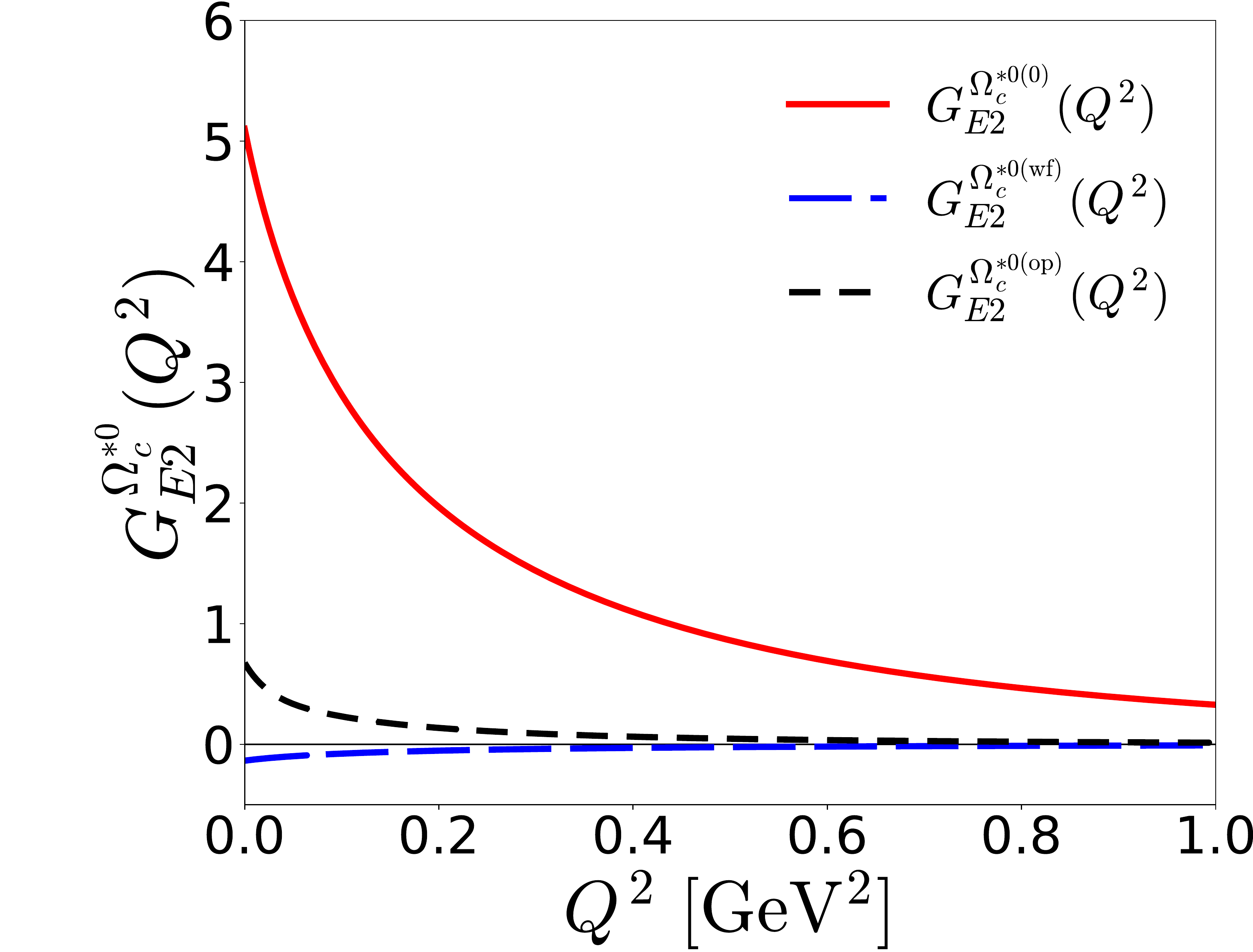}
\caption{Linear $m_{\mathrm{s}}$ corrections from the current-quark
  mass term in the effective chiral action
  $G_{E2}^{B_c^*(\mathrm{op})}$ and from the collective wavefunctions
  $G_{E2}^{B_c^*(\mathrm{wf})}$, which are drawn respectively in the
  short-dashed and long-dashed curves.  }   
\label{fig:3}
\end{figure}

In Fig.~\ref{fig:2}, we show how much the effects of flavor SU(3)
symmetry breaking contribute to the $E2$ form factors of the baryon
sextet with spin 3/2. As expressed in Eqs.~\eqref{eq:ms_op}
and~\eqref{eq:ms_wf}, there are two different $m_{\mathrm{s}}$
corrections to the $E2$ form
factors. The first one ${\cal{G}}^{B_{6}(\text{op})}_{E2}(Q^2)$ arises
from the current-quark mass term in the effective chiral action given 
in Eq.~\eqref{eq:echl}, whereas the second one comes from the
wavefunction corrections~\eqref{eq:mixedWF1}. each correction affects
$E2$ form factors in a different way, as shown in
Fig.~\ref{fig:3}. The wavefunction corrections to the $E2$ form factor
of $\Sigma_{c}^{*++}$ are negligibly tiny and the corrections from
the current-quark mass term is also small. As a result, the
$m_{\mathrm{s}}$ corrections turn out to be negligible, as shown in
the upper left panel of Fig.~\ref{fig:2}. On the other hand, the
wavefunction corrections contribute noticeably to the $E2$ form
factors of $\Sigma_c^{*+}$, while those from
the current-quark mass term are of the same order as in the case of
$\Sigma_{c}^{*++}$. In the case of $\Sigma_c^{*0}$ and $\Xi_c^{*0}$,
the wavefunction corrections to $G_{E2}^{\Sigma_c^{*0},\Xi_c^{*0}}$ are
even larger than those from the mass term. This can be understood 
by examining Eqs.~\eqref{eq:ms_op} and~\eqref{eq:ms_wf}. 

\begin{figure}[htp]
\centering
\includegraphics[scale=0.26]{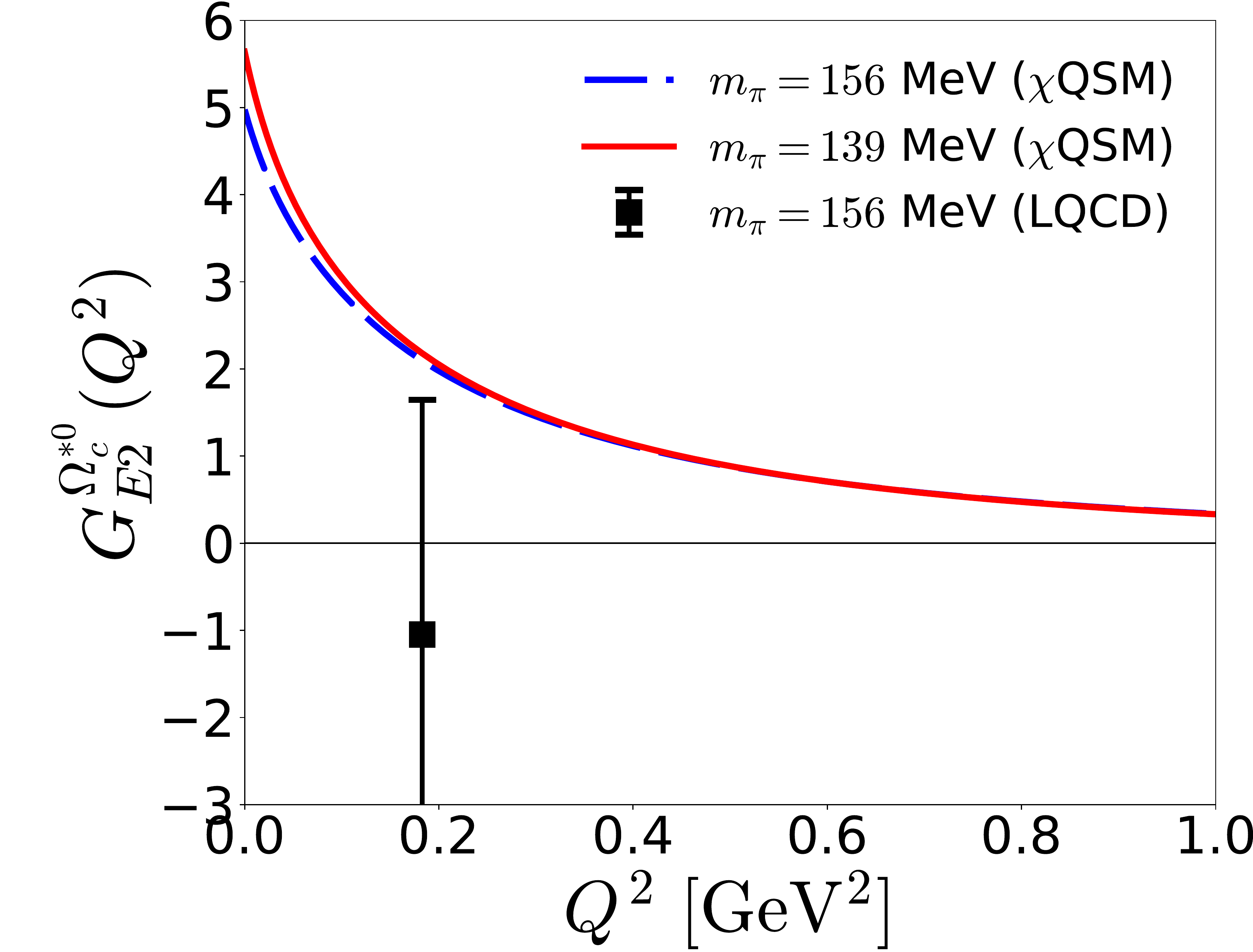}
\includegraphics[scale=0.26]{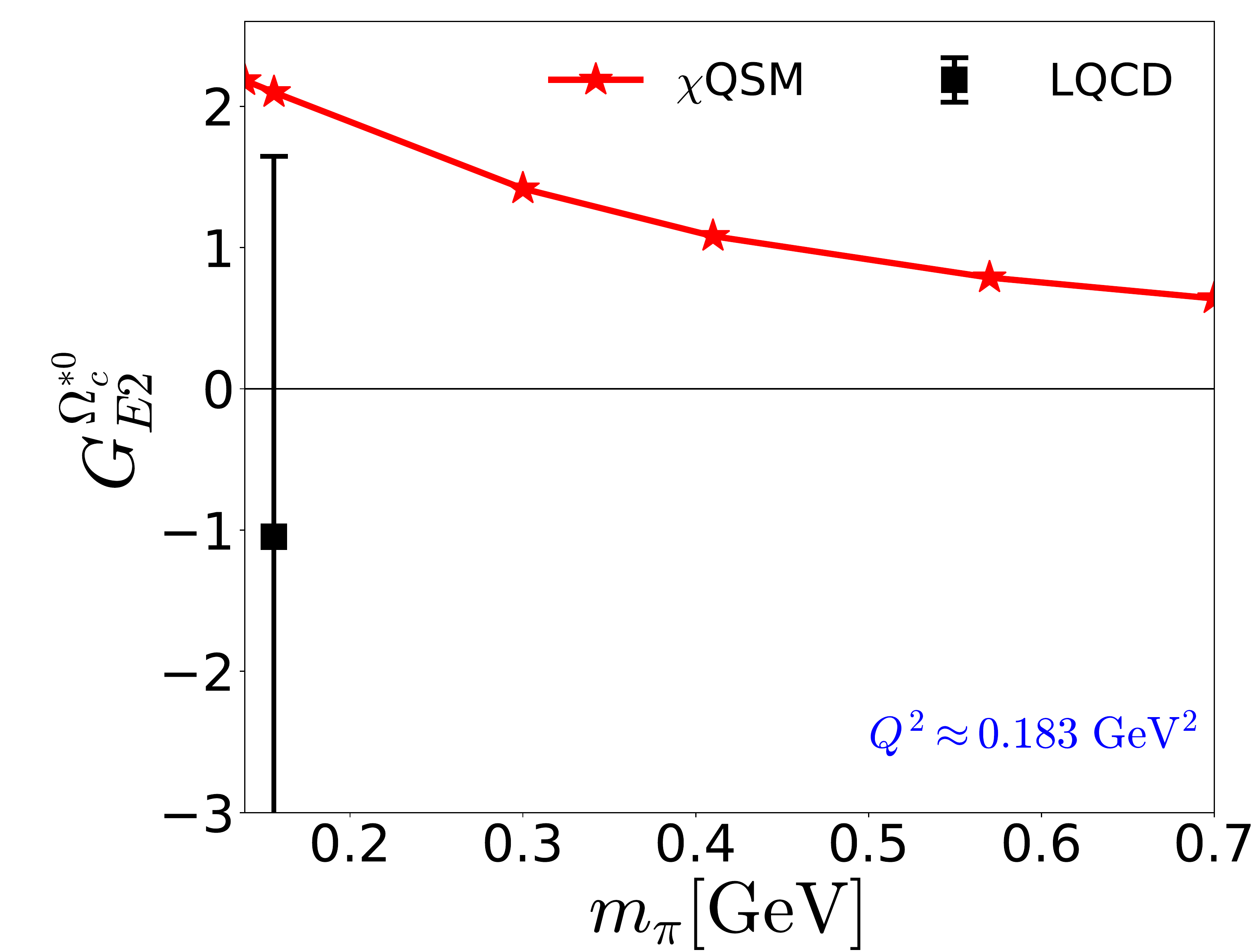}
\caption{Electric quadrupole form factors of the baryon sextet with
  spin 3/2 in comparison with the data from the lattice QCD. The data
  of the lattice  QCD is taken from Ref.~\cite{Can:2015exa}.} 
\label{fig:4}
\end{figure}
In the left panel of Fig.~\ref{fig:4}, we compare the results for the
$E2$ form factors of the $\Omega_c^{*0}$ baryon with that from the
lattice calculation. We employ for this comparison the unphysical pion
mass $m_\pi=156$ MeV that is used in the lattice calculation. 
Note that there is only one lattice data with large uncertainty. 

We anticipate more accurate lattice data in the
near future, so that one can draw a clear conclusion. 
In the right panel of Fig.~\ref{fig:4}, we depict the results of
$G_{E2}^{\Omega_c^{*0}}$ as a function of the pion mass $m_\pi$ with
$Q^2=0.183\,\mathrm{GeV}^2$ fixed. As expected, the present results
fall off slowly as $m_\pi$ increases. 

\setlength{\tabcolsep}{5pt}
\renewcommand{\arraystretch}{1.5}
\begin{table}[htp]
\centering
\caption{Electric quadrupole moments of the baryon sextet.} 
\label{tab:1}
 \begin{tabular}{ c | c c c c c c } 
  \hline 
    \hline 
 $\mathcal{Q}_{B}$ [$e\cdot$ fm$^{2}$] & $\Sigma^{*++}_{c}$  &
 $\Sigma^{*+}_{c}$   & $\Sigma^{*0}_{c}$ & 
 $\Xi^{*+}_{c}$  &  $\Xi^{*0}_{c}$ &  $\Omega^{*0}_{c}$ \\
  \hline 
$m_{s} = 180$~MeV&$-0.0490 $ & $-0.0058$  &  $0.0373$ & $-0.0234$
& $0.0330$  &  $0.0286$  \\  
$m_{s} = 0$~MeV&$-0.0518$ & $-0.0129$  &  $0.0259$ & $-0.0129$
& $0.0259$  &  $0.0259$ \\  
\hline
 \hline
\end{tabular}
\end{table}
For completeness, we present the results for the electric quadrupole
moments of  the baryon sextet with spin 3/2. Table~\ref{tab:1} lists
those of the $\mathcal{Q}_B$ in the second and third rows, which
correspond to the SU(3) symmetric and breaking cases, respectively.  
As already shown in Fig.~\ref{fig:2}, those of the charged baryon
sextet have negative values of $\mathcal{Q}_{B}$, which indicates that
the positively charged singly heavy baryons with spin 3/2 take oblate
shapes. On the other hand, those of the neutral ones get positive
values, so they are distorted in prolate forms. It is interesting to
see that the $\mathcal{Q}_B$ of the doubly positive-charged
$\Sigma_c^*$ is approximately 8 times larger than that of the singly
positive-charged one. This can be understood by examining
Eqs.~\eqref{eq:E2final}.  
\section{Summary and conclusion}
In the present work, we have investigated the
electric quadrupole form factors of the lowest-lying singly heavy
baryons with spin 3/2 in a pion mean-field approach, also knwon as the
SU(3) chiral quark-soliton model. In the limit of an infinitely heavy
quark, a heavy quark inside a singly heavy baryon can be regarded as a
mere static color source. This means that the $N_c-1$ light valence
quarks govern the quark dynamics inside a heavy baryon. The presence
of the $N_c-1$ light valence quarks make the vacuum polarized, which
produces the pion mean fields. The $N_c-1$ valence quarks are bound by
the attraction provided by the pion mean fields self-consistently,
from which a  colored soliton consisting of the $N_c-1$ valence quarks 
arises. We call this soliton as a $N_c-1$ soliton. Then the singly
heavy baryon can be constructed by coupling the $N_c-1$ soliton with a
heavy quark. This is called the pion mean-field approach for the
singly heavy baryons. Based on this pion mean-field approach, we 
computed the electric qudrupole form factors of the baryon sextet with
spin 3/2, taking into account the rotational $1/N_c$ and linear
$m_{\mathrm{s}}$ corrections. 

We first examined the valence- and sea-quark contributions separately.
As in the case of the baryon decuplet, the contributions from the sea
quarks or the Dirac-sea level quarks govern the electric quadrupole 
form factors, in particular, in the smaller $Q^2$ region. 
Considering the fact that the electric quadrupole moment of a baryon
provides information on how the baryon is deformed, we can draw the
following physical implications: the deformation of a singly heavy
baryon is also mainly governed by the sea-quark contributions or the
pion cloud effects. We found a similar feature in the case of the
baryon decuplet. The effects of the explicit flavor SU(3) symmetry
breaking are also sizable except for the case of the $\Sigma_c^{*++}$
and $\Omega_c^{*0}$. Since there are two different linear
$m_{\mathrm{s}}$ corrections, we have scrutinized each effect in
detail. To compare the present results with those from the lattice
calculation, we have computed the electric quadrupole form factor with
the unphysical value $m_\pi=156$ MeV adopted, which was used by the
lattice work. We also showed how the value of the form factor at a
fixed $Q^2$ is changed as the $m_\pi$ increases. As expected from
previous works, the value of the form factor falls off as $m_\pi$
increases. We also presented the results for the electric quadrupole
moment. The charged singly-heavy baryons have consistently negative
values of the electric quadrupole moments. This indicates that the
charged baryons take oblate shapes. On the other hand, the neutral
baryons take prolate shapes, having the positive values of the electric
quadrupole moments.

\begin{acknowledgments}
The authors are grateful to Gh.-S. Yang for valuable discussions. 
They want to express the gratitude to M. Oka and K. U. Can for
providing us with the lattice data. The present work was supported by
Basic Science Research Program through the National Research
Foundation of Korea funded by the Ministry of Education, Science and
Technology (2018R1A2B2001752 and 2018R1A5A1025563).  J.-Y. Kim
acknowledges partial support by the DAAD doctoral scholarship.  
\end{acknowledgments}

\begin{appendix}
\section{Densities for the $E2$ form factor and moments of
  inertia\label{app:A}}   
In this Appendix, we provide the explicit expressions for the
$\mathcal{I}_{1E2}$ and $\mathcal{K}_{1E2}$ densities of the electric
quadrupole form factors in Eq.~\eqref{eq:magden}
\begin{align}
\mathcal{I}_{1E2}(\bm{z})&= -\frac{(N_c-1)}{2\sqrt{10}} \sum_{n \ne
\mathrm{val} }\frac{1}{E_{n}-E_{\mathrm{val}}}{\langle\mathrm{val}
| \bm{\tau} | n\rangle} \cdot{\langle n |\bm{z} \rangle
\{ \sqrt{4\pi}Y_{2}  \otimes\tau_{1}  \}_{1}\langle \bm{z} |
\mathrm{val}\rangle} \cr 
&  + \frac{N_c}{4\sqrt{10}} \sum_{n,m} {\cal{R}}_{3}(E_n,E_m)
  {\langle n |   \bm{\tau} | m \rangle} \cdot{\langle m | \bm{z}
  \rangle \{   \sqrt{4\pi} Y_{2}  \otimes \tau_{1}  \}_{1} \langle
  \bm{z} | n   \rangle}  ,\cr 
 \mathcal{K}_{1E2}(\bm{z})&=
  -\frac{(N_c-1)}{2\sqrt{10}}\sum_{n  \ne 
 \mathrm{val} } \frac{1}{E_{n}-E_{\mathrm{val}}} {\langle\mathrm{val}
 |  \gamma^{0} \bm{\tau} | n \rangle} \cdot {\langle n | \bm{z} \rangle
 \{ \sqrt{4\pi} Y_{2}  \otimes \tau_{1}  \}_{1} \langle \bm{z} |
 \mathrm{val} \rangle} \cr  
& -  \frac{N_c}{4\sqrt{10}}\sum_{n,m} {\cal{R}}_{5}(E_n,E_m) {\langle
  n | \gamma^{0} \bm{\tau} | m \rangle} \cdot {\langle m | \bm{z}
  \rangle  \{ \sqrt{4\pi} Y_{2}  \otimes \tau_{1}  \}_{1} \langle
  \bm{z} | n   \rangle}, 
\end{align}
where the regularization functions are defined by 
\begin{align}
&{\cal{R}}_{3}(E_{n},E_{m}) = \frac{1}{2 \sqrt{\pi}} \int^{\infty}_{0}
  \phi(u) \frac{du}{\sqrt{u}} \left[ \frac{ e^{-u E_{m}^{2}}- e^{-u
  E_{n}^{2}}}{u(E^{2}_{n} - E^{2}_{m})} -\frac{E_{m} e^{-u
  E_{m}^{2}}+E_{n} e^{-u E_{n}^{2}}}{E_{n} + E_{m}}  \right ], \cr 
&{\cal{R}}_{5}(E_{n},E_{m}) =
  \frac{\mathrm{sign}(E_{n})-\mathrm{sign}(E_{m})}{2(E_{n}-E_{m})},
\end{align}
with the proper-time regulator $\phi(u)$~\cite{Christov:1995vm}.
Here, $|\mathrm{val}\rangle$ and $|n\rangle$ denotes the state of the 
valence and sea quarks with the corresponding eigenenergies
$E_{\mathrm{val}}$ and $E_n$ of the single-quark Hamiltonian $h(U_c)$,
respectively.  

The moments of inertia ($I_{1}, I_{2}$) and anomalous moments of
inertia ($K_{1}, K_{2}$) are expressed respectively as 
\begin{align}
I_{1} &= \frac{(N_{c}-1)}{6} \sum_{n\neq\mathrm{val}}
        \frac{1}{E_{n}-E_{\mathrm{val}}}\langle \mathrm{val} |
        \bm{\tau} | n \rangle \cdot \langle n | \bm{\tau} |
        \mathrm{val} \rangle + \frac{N_{c}}{12}\sum_{n,m\neq n}\langle
        m | \bm{\tau} | n \rangle \cdot \langle n | \bm{\tau} | m
        \rangle \mathcal{R}_{3}(E_n,E_m), \cr 
I_{2} &= \frac{(N_{c}-1)}{4} \sum_{n^{0}}
        \frac{1}{E_{n^{0}}-E_{\mathrm{val}}}\langle \mathrm{val} |
        n^{0} \rangle  \langle n^{0}  | \mathrm{val} \rangle
        +\frac{N_{c}}{4} \sum_{n^{0},m}\langle m | \bm{\tau} | n^{0}
        \rangle \langle n^{0}  | m \rangle
        \mathcal{R}_{3}(E_{n^{0}},E_m), \cr 
K_{1} &= \frac{(N_{c}-1)}{6} \sum_{n\neq\mathrm{val}}
        \frac{1}{E_{n}-E_{\mathrm{val}}}\langle \mathrm{val} |
        \bm{\tau} | n \rangle \cdot \langle n | \gamma^{0} \bm{\tau} |
        \mathrm{val} \rangle + \frac{N_{c}}{12}\sum_{n,m\neq n}\langle
        m | \bm{\tau} | n \rangle \cdot \langle n | \gamma^{0}
        \bm{\tau} | m \rangle \mathcal{R}_{5}(E_n,E_m), \cr 
K_{2} &= \frac{(N_{c}-1)}{4} \sum_{n^{0}}
        \frac{1}{E_{n^{0}}-E_{\mathrm{val}}}\langle \mathrm{val} |
        n^{0} \rangle  \langle n^{0} |\gamma^{0}  | \mathrm{val}
        \rangle +\frac{N_{c}}{4}\sum_{n^{0},m}\langle m  | \bm{\tau} |
        n^{0} \rangle \langle n^{0} |\gamma^{0} | m \rangle
        \mathcal{R}_{5}(E_{n^{0}},E_m). 
\end{align}

\section{Fixing the model parameters\label{app:B}}  
The chiral condensate and the pion decay constant can be derive from
the effective chiral action given in Eq.~\eqref{eq:echl}. The chiral
condensate are written as 
\begin{align}
\langle \overline{\psi}\psi \rangle = - \int
  \frac{d^{4}p_{E}}{(2\pi)^{4}}
  \frac{8N_{c}M}{p^{2}_{E}+M^{2}}\bigg{|}_{\mathrm{reg}} = M
  \frac{N_{c}}{2\pi^{2}} \int^{\infty}_{0}\phi(u)
  \frac{du}{u^{2}}e^{-u M^{2}}, 
\label{eq:1}
\end{align}
and the pion decay constant are given by 
\begin{align}
f^{2}_{\pi} = - \int \frac{d^{4}p_{E}}{(2\pi)^{4}}
  \frac{4N_{c}M^{2}}{(p^{2}_{E}+M^{2})^{2}}\bigg{|}_{\mathrm{reg}} =
  M^{2} \frac{N_{c}}{4\pi^{2}} \int^{\infty}_{0}\phi(u)
  \frac{du}{u}e^{-u M^{2}}, 
\label{eq:2}
\end{align}
with proper-time regulator $\phi= c \theta(u-\Lambda^{-2}_{1})+ (1-c)
\theta(u-\Lambda^{-2}_{2})$. 
The pion mass is determined by the pole position of the pion
propagator that is obtained by a low-energy effective chiral theory
given by Eq.~\eqref{eq:echl} 
\begin{align}
m^{2}_{\pi} = \frac{\overline{m} \langle \overline{\psi} \psi
  \rangle}{f^{2}_{\pi}} + \mathcal{O}(\overline{m}^{2}). 
\label{eq:3}
\end{align}
The above-given expressions satisfy the Gell-Mann–Oakes–Renner
(GMOR) relation. With Eqs.~\eqref{eq:1},~\eqref{eq:2} and
~\eqref{eq:3}, one can determine the cutoff mass. The average value of
the up and down current quark masses is obtained as 
$\overline{m}=6.13~\mathrm{MeV}$. The strange current quark mass
$m_{s}$ is fixed by the hyperon mass 
splittings by treating $m_{s}$ perturbatively up to the second-order
corrections~~\cite{Blotz:1992pw, Christov:1995vm, Kim:2018xlc}. The
preferable value of $m_{s}$ is found to be $m_{s}=180~\mathrm{MeV}$. 
\end{appendix}


\end{document}